\DeclareUnicodeCharacter{2061}{}
\documentclass[final,3p,times,authoryear]{elsarticle}
\usepackage{amssymb}
\usepackage{amsthm}
\usepackage{amsmath}

\journal{New Astronomy}

\begin{document}

\begin{frontmatter}



\title{Orbital Dynamics of Atlas (S XV): its Current Orbit and the Recent Past}

 \author[label1]{Demétrio Tadeu Ceccatto}
 \author[label1]{Nelson Callegari Jr.}
 \author[label2,label3]{Gabriel Teixeira Guimarães}
 \author[label4]{Karyna Gimenez}
 \affiliation[label1]{organization={São Paulo State University (UNESP), Institue of Geosciences and Exact Sciences, Rio Claro},
             addressline={Av. 24A, 1515},
             city={Rio Claro},
             postcode={13506-900},
             state={São Paulo},
             country={Brazil}}
 \affiliation[label2]{organization={Graduate University for Advanced Studies, SOKENDAI Shonankokusaimura, Hayama},
             addressline={Miura District Kanagawa},
             city={Tokyo},
             postcode={240-0115},
             country={Japan}}
\affiliation[label3]{organization={Division of Science, National Astronomical Observatory of Japan, NAOJ},
            addressline={Osawa, Mitaka},
            city={Tokyo},
            postcode={181-8588},
            state={},
            country={Japan}}
 \affiliation[label4]{organization={Observatório do Valongo, Universidade Federal do Rio de Janeiro (UFRJ)},
             addressline={Ladeira do Pedro Antônio 43},
             city={Rio de Janeiro},
             postcode={20080-090},
             state={Rio de Janeiro},
             country={Brazil}}

%%\author[first]{Demétrio Tadeu Ceccatto}
%%\affiliation[first]{organization={São Paulo State University (UNESP), Institue of Geosciences and Exact Sciences},%Department and Organization
%%            addressline={Av. 24A, 1515}, 
%%            city={Rio Claro},
%%            postcode={13506-900}, 
%%            state={São Paulo},
%%            country={Brazil}}

\begin{abstract}
%The \textit{Cassini-Huygens} mission contributed significantly to the understanding of the dynamic complexity that involves the orbits of Saturn's natural satellites. New ephemeris was determined with a higher level of precision, making it possible to study in detail the secular and resonant dynamics and, in particular, the effects of the 54:53 Prometheus-Atlas and 70:67 Pandora-Atlas mean motion resonances. In this work, we performed a detailed analysis of the current orbit of Atlas under different scenarios identifying secular, resonant perturbations due to Prometheus and Pandora, and also due to Saturn's oblateness. We mapped the domain of the 54:53 Prometheus-Atlas and 70:67 Pandora-Atlas resonances in the phase space considering a dense set of initial conditions of semi-major axis and eccentricity for several clones of Atlas. \textbf{Our maps allowed us to identify several 1st and 2nd-order resonances in phase space.} The combined analysis of the mappings of the 54:53 Prometheus-Atlas and 70:67 Pandora-Atlas resonances revealed the overlap of both resonant domains in phase space. We have therefore identified a physical reason for the chaotic motion of Atlas previously detected ($\sim$ 0.1 year$^{-1}$). Finally, we investigate the orbital dynamics of Atlas considering Prometheus in a past configuration dictated by tidal forces in Saturn, which reveals that Atlas-Prometheus could be a co-orbital system at 1.8×$10^8$ years.
This study comprehensively analyzes Atlas's current orbit, focusing on the secular and resonant perturbations caused by Prometheus, Pandora, and Saturn's oblateness. We performed numerical integration of the exact equations of motion for a dense ensemble of Atlas clone satellites. Through spectral analysis and interpretation of these orbits on dynamical maps, we identified the domain of the 54:53 Prometheus-Atlas and 70:67 Pandora-Atlas mean-motion resonances, showing that Atlas lies on the boundary of the separatrices of each of these resonances. We also identified the domains for the multiplets $\Psi_{1}$, $\Psi_{2}$, $\Psi_{3}$ and $\Psi_{4}$ associated with 70:67 resonance. Additionally, we explored the variation in Prometheus's eccentricity, demonstrating that as eccentricity increases (or decreases) in the  54:53 resonance domain correspondingly decreases (or increases). This combined analysis, between the above mappings, revealed qualitatively the overlap between the 54:53 and 70:67 resonances, which are responsible for the chaotic behavior of Atlas's orbit. We quantified chaotic motion in frequency space and found that the vicinity of Atlas is characterized by weak to moderate chaos, rather than strong chaos. Finally, we investigated Atlas's recent past, considering Prometheus's migration under the influence of Saturn's tidal forces. This analysis reveals several resonances crossed in the past, particularly focusing on the Atlas-Prometheus pair, which exhibited a co-orbital configuration.

\end{abstract}




\end{frontmatter}

%\tableofcontents



\section{Introduction}
\label{1}

Saturn’s natural satellites display a variety of orbital configurations, with Atlas being a notable example. Atlas is a small satellite with an average radius of approximately 14.9 km (see Thomas and Helfenstein 2020). Its orbit is gravitationally influenced by Prometheus and Pandora, which have average radii of approximately 42.8 km and 40 km, respectively. By combining data from various sources, including ground-based telescopes, the \textit{Hubble Space Telescope}, and spacecraft observations, Spitale et al. (2006) refined the ephemeris for Atlas, Prometheus, and Pandora. They suggested that Atlas’s orbital dynamics are primarily affected by the 54:53 mean-motion resonance (MMR) with Prometheus and, to a lesser extent, by the 70:67 resonance with Pandora. Spitale et al. (2006) also made the first conjecture that Atlas's orbit might be chaotic, a fact later confirmed by Cooper et al. (2015). Using the Fast Lyapunov Indicator tool, they determined a Lyapunov exponent of 0.1 year$^{-1}$. Cooper et al. (2015) also provided the updated mass values for Atlas, Prometheus, and Pandora. Renner et al. (2016) further studied the chaotic dynamics of Atlas using a truncated model with two degrees of freedom, concluding that Prometheus is the primary source of disturbances in Atlas's orbit. Ceccatto et al. (2022) mapped the orbital vicinity of Atlas in the frequency domain and found that Atlas's current orbit is located on the boundary of the 54:53 resonance with Prometheus. Inspired by the short Lyapunov time, $\sim$ 10 years, Pereira et al. (2024) analyzed the temporal evolution of Atlas's current orbit. They concluded that initial displacement primarily affected the angular components of the Atlas clones, introducing the chaos, but did not significantly influence the radial component. This behavior, they proposed, confines the chaotic nature of Atlas's orbit.

Our primary objective in this study is to explore and extend the key findings of previous works (e.g. Spitale et al. 2006; Cooper et al. 2015; Renner et al. 2016; Ceccatto et al. 2022; Pereira et al. 2024).

We present a collection of dynamical maps constructed through tens of thousands of numerical simulations for several dense sets of initial conditions of Atlas clones. These maps help us identify regions of the phase space where orbital motion remains regular, allowing us to pinpoint the domain of the resonances. For the construction of these maps, we use two different models. The first model involves mapping the phase space in the frequency domain (see Laskar 1993; Michtchenko and Ferraz-Mello 2001; Callegari and Yokoyama 2010a).  The second model utilizes the maximum variation for the semi-major axis, defined as $\Delta{a}=a_{max}-a_{min}$ (see Rodríguez and Callegari 2021). Our methodology consists of numerically solving the exact equations of motion over extended time intervals for the orbital motion of Atlas clones, which orbit a non-spherical body and are disturbed by other satellites. The spectra of these orbits were obtained and analyzed numerically (see Section~\ref{2}). 

In Section~\ref{3.1}, we investigate the current orbit of Atlas and its vicinity. Section~\ref{3.2} addresses the secular variations in eccentricity and inclination, caused by the combined effects of Prometheus, and Saturn's oblateness on Atlas's orbit. Section~\ref{4.1} explores how Prometheus's orbital eccentricity influences Atlas's orbit, as Cooper et al. (2015) suggest that it is a key factor in Atlas's stability. In Section~\ref{4.2}, we examine Pandora's contributions to the Atlas's orbit and, by interpreting the dynamical maps, we quantitatively demonstrate the overlap between the 54:53 Prometheus-Atlas and 70:67 Pandora-Atlas resonances. In Section~\ref{5} we measure the chaos in Atlas's orbit using the Laskar diffusion coefficient (Laskar 1993). 

In Section~\ref{6}, investigates Atlas's recent dynamical past through numerical analysis, considering the tidal evolution of Prometheu's orbit. Using the methodology described in Giuppone et al. (2022), Section~\ref{6.5} shows that the orbits of Atlas and Prometheus could have been co-orbital approximately $10^8$ years ago. Finally, Section~\ref{7} presents the main conclusions of this work.

\section{Methodology}
\label{2}

%In this work, we follow the methodology given in Callegari and Yokoyama (2010a, 2020), and Callegari et al. (2021) where we numerically integrate the equations of motion for a system formed by \textit{N} mutually perturbed satellites orbiting around Saturn \textbf{and we study the fundamental frequencies of these orbital elements using the methodology given in Laskar (1990).}

In this study, we followed the methodology outlined by Callegari and Yokoyama (2010a, 2020) and Callegari et al. (2021) where the authors numerically integrate the equations of motion for a system formed by \textit{N} mutually perturbed satellites orbiting Saturn. The chaos detection method used by Callegari and Yokoyama (2010a, 2020) and Callegari et al. (2021) was initially introduced by Michtchenko and Ferraz-Mello (2001). Additionally, the characterization of the phase space in the frequency domain was first presented by Laskar (1990) and was further applied by Michtchenko and Ferraz-Mello (2001).

We utilized two distinct numerical models: i) the system of exact equations given by Equations 1 to 5 in Callegari and Yokoyama (2010a), where the equations of motion are integrated under the influence of the terms $J_2$ and $J_4$ (coefficients of Saturn's potential expansion); ii) direct application of the MERCURY package (Chambers 1999) with the addition of the term {$J_6$}. Although the perturbative effects of $J_6$ are small, compared to those of $J_2$ and $J_4$, we still accounted for them. In both cases i) and ii) we used the Everhart “RA15” code to solve systems of ordinary differential equations (Everhart 1985).

The physical parameters of Saturn and the initial conditions of Atlas, Prometheus, and Pandora are listed in Tables~\ref{table 1} and ~\ref{table 2}, respectively. The masses of the satellites were obtained from Thomas and Helfenstein (2020), and the initial osculating elements were derived from the \textit{Horizons} ephemeris system (http://ssd.jpl.nasa.gov/horizons.cgi).

\begin{table}[h!]
	\caption{Physical constants for Saturn: \textit{Horizons}$^a$, Jacobson et al. (2006)$^b$}
	   \begin{center}
		  \begin{tabular}{cc}
		  	\hline
                 \hline
				Constant & Numerical value   \\
				\hline
				$GM^a$ (Km$^3 s^{-2}$)& 3.7931208x$10^7$ \\
				%	\hline
				Equatorial radius$^a$ (Km) & 60268$\pm$4 \\
				%	\hline
				$J_2^{b}$& 16290.71x$10^{-6}$ \\
				%	\hline
				$J_4^{b}$ & -935.83x$10^{-6}$ \\
				%	\hline
				$J_6^{b}$& 86.14x$10^{-6}$ \\
			 \hline
		\end{tabular}
	\end{center}
 \label{table 1}
\end{table}

	\begin{table}[h!]
		\caption{The Ephemeris \textit{Horizons} data on January 1st, 2000, had shown osculating elements for Atlas, Prometheus, and Pandora (Accessed on 05/29/2022). The masses have been taken from Thomas and Helfenstein (2020).}
       \begin{center}		    
		     \begin{tabular}{cccccccc}
			\hline
            \hline
			 & Atlas & Prometheus & Pandora \\
   		\hline
			\textit{Mass} (kg) & 0.575x$10^{16}$ & 15.6x$10^{16}$ & 13.7x$10^{16}$ \\
            \textit{a} (km) & 138,325.32 & 140,024.64 & 142,346.13 \\
            \textit{e} & 0.0059 & 0.0025 & 0.0015 \\
            $i$ (degree) & 0.0042 & 0.0081 & 0.050 \\
			$\omega$ (degree) & 200.77 & 201.36 & 332.15 \\
			$\Omega$ (degree) & 235.44 & 309.14 & 149.98 \\
            $n$ (degree/day) & 592.97 & 581.87 & 567.69 \\
            $\ell$ (degree/day) & 357.64 & 115.18 & 288.84 \\
            \hline
            \end{tabular}
        \end{center}
        \label{table 2}
	\end{table}

We analyzed the spectra of Atlas's orbital element clones using the FFT algorithm provided by Press et al. (1996). To quantify the spectra, we assigned a spectral number, \textbf{N}, representing the number of peaks in the spectrum that exceed a predetermined percentage of the highest peak. This percentage, referred to as the reference amplitude (RA), was generally set at  0,1\% or 5\% of the highest peak value. Additionally, we excluded peaks related to short-periods of less than 15 days from the count, ensuring that the dynamical maps remain free from short-term variations. 

The dynamical maps are constructed using a color palette, with different tones representing various values of \textbf{N}. We set \textbf{N}*=100 as the cutoff value, meaning the same color is assigned when \textbf{N} $>$ \textbf{N}*. Only the initial conditions of the grid, typically the initial eccentricity versus the initial semi-major ($a_0$, $e_0$), are changed. All other initial elements of Atlas clones and other satellites were based on data from January 1st, 2000.

In this work, we primarily adopted the osculating elements for frequency analysis, which includes the construction of the dynamical maps and IPS. However, we used geometric elements when analyzing the current orbits of Atlas clones obtained using model (ii). In this context, the osculating orbital elements can experience significant short-period variations due to $J_2$, while the corresponding geometric elements remain largely unaffected by these variations (see Figures~\ref{fig_5}(a,b)). The geometric elements were calculated from the {state vector}\footnote{The vector that contains the position coordinates and the respective instantaneous velocities for the satellite or the particle obtained from the osculating elements of the numerical simulation.} obtained from the osculating orbit for Atlas clones in our simulations, using the algorithm developed by Renner and Sicardy (2006). The most affected elements are the eccentricity and argument of the pericenter. A detailed discussion on this can be found in Callegari and Yokoyama (2020) and Callegari et al. (2021), particularly for satellites orbiting near a strong $J_2$ field.

\section{Conservative dynamics I: The domain of 54:53 Prometheus-Atlas mean-motion resonance in the phase spaces}
\label{3}

\begin{figure}[!ht]
	\centering 
	\includegraphics[width=1.0 \columnwidth,angle=0, scale = 0.75]{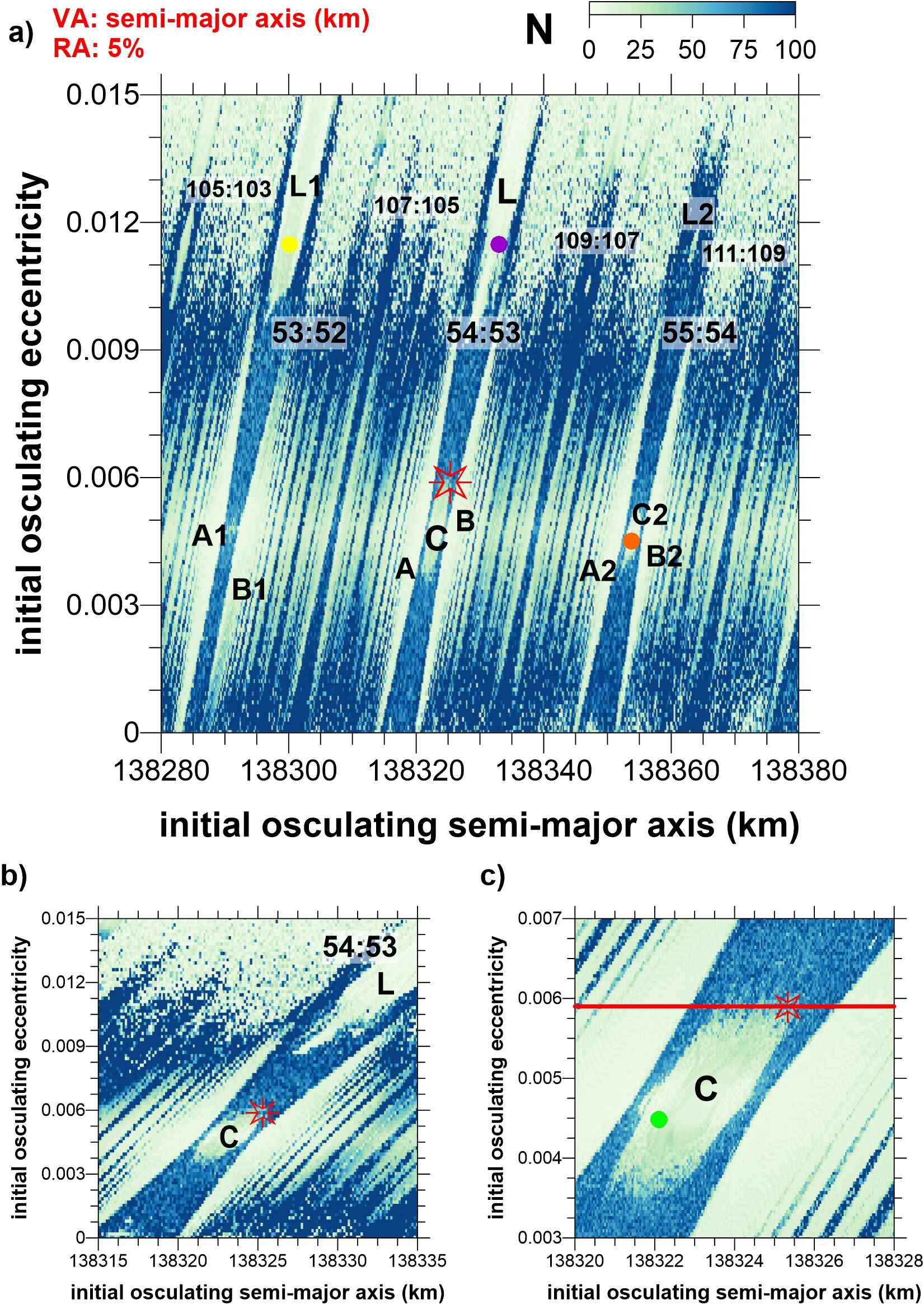}	
	\caption{(a) Dynamical map in the vicinity of Atlas' current orbit considering the initial date January 1st, 2000, where $a_0$ $\sim$ 138,325.32 km and $e_0$ $\sim$ 0.0059 (represented by the red star). \textbf{C} and \textbf{L} are the respective Corotation and Lindblad zones. For each initial condition, the total integration time for the numerical simulation is 188,743.68 days ($\sim$516.75 years) sampled every 0.18 days. \textbf{N} is the spectral number. In this map, we have 51,005 orbits of Atlas clone satellites, with $\delta$$a$=0.2 km, $\delta$$e$=0.00015 and RA=5\% using the semi-major axis as the variable of analysis (VA). For Atlas's orbit, we have the 53:52 and 55:54 resonances on the left and right, respectively, and several second-order resonances are identified. Colored circles and red star represent initial conditions for the orbits seen in Figure~\ref{fig_2}. \textbf{A}, \textbf{A1}, \textbf{A2}, \textbf{B}, \textbf{B1}, \textbf{B2} are regular motion regions not associated with any resonances and \textbf{L1} represent the Lindblad zone associated with 53:52 resonance and \textbf{C2} and \textbf{L2} the respective zones for Corotation and Lindblad associated with resonance 55:53. (b) the same (a) but considering the domain for the 54:53 resonance. (c) A detailed representation of the Corotation zone (C).} 
	\label{fig_1}%
\end{figure}

In this section, we investigate the domains of the 54:53 resonance in the vicinity of Atlas's current orbit, with {Prometheus as the primary perturber}\footnote{The Mimas-Tethys pair of satellites were not included in our simulations because previous tests showed that perturbations from this pair do not alter the phase space seen in Figure~\ref{fig_1}.}.

Dynamical maps are, particularly, useful for identifying regions in phase space that hold significant physical relevance. Several such regions are shown in Figure~\ref{fig_1}, where the semi-major axis serves as the variable of analysis (VA). The intervals used in constructing the map were $\Delta{a}$= [138,280 km, 138,380 km] and $\Delta{e}$= [0, 0.015]. The red star in Figure~\ref{fig_1} indicates Atlas's location on the adopted date, January 1st, 2000, with initial values for the semi-major axis and eccentricity being approximately $a_0$ $\sim$ 138,325.32 km, $e_0$ $\sim$ 0.0059, respectively (see Table~\ref{table 2}). The map in Figure~\ref{fig_1}(a) was generated following the numerical integration and Fourier analysis of 51,005 orbits.

\begin{figure}[!ht]
	\centering 
	\includegraphics[width=1.0 \columnwidth,angle=0, scale = 0.85]{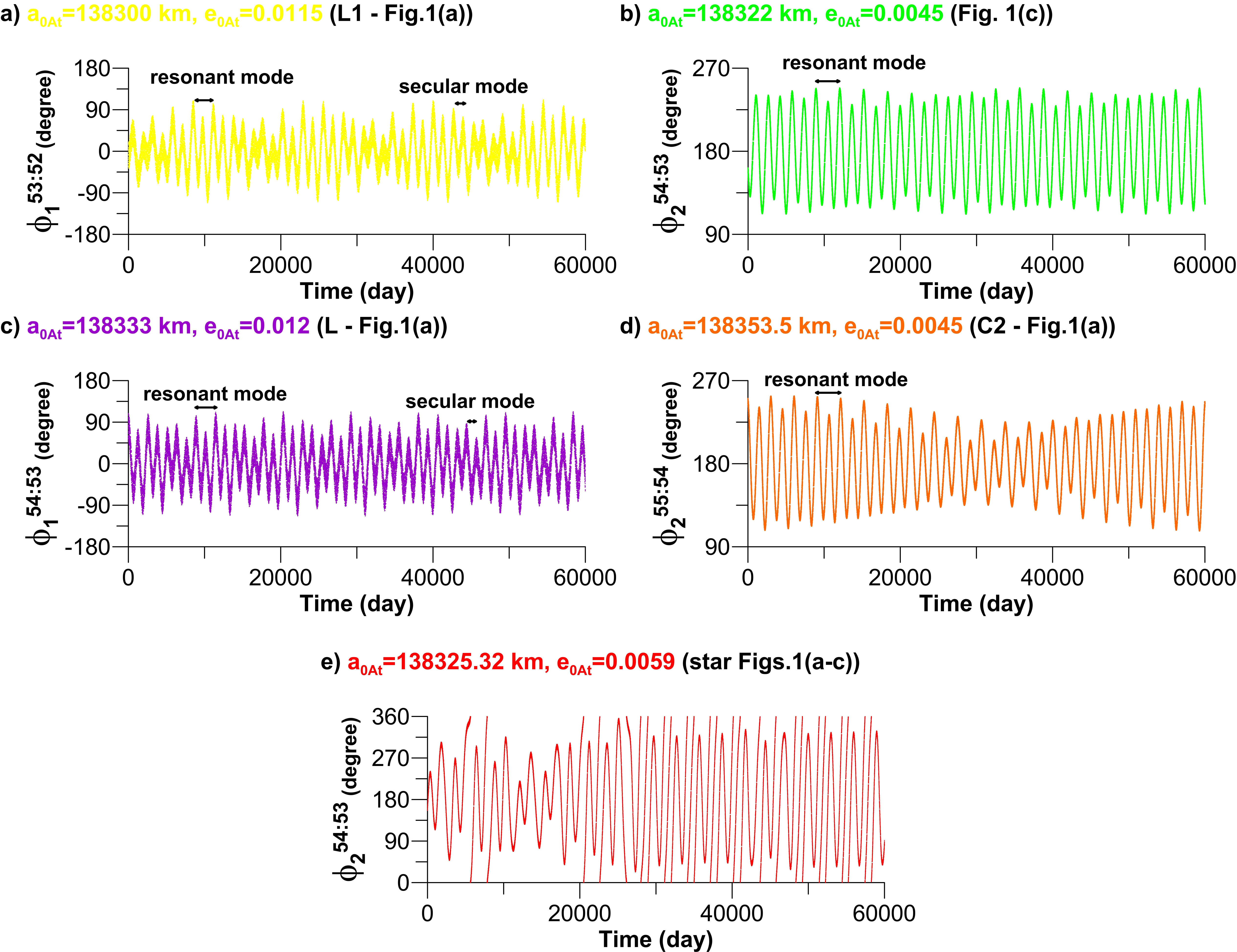}	
	\caption{Temporal variation of the angles $\phi_1^{m+1:m}$ and $\phi_2^{m+1:m}$, where $m$ = 52, 53 and 54, associated with Lindbald and Corotation resonances, respectively. For the Atlas clone orbits with initial conditions given in Figures~\ref{fig_1}(a,c) (colored circles): (a) and (c) the oscillate of the Lindblad angle around 0 and in (b) and (d) the oscillate of the Corotation angle around 180°. (e) Alternation between oscillation and circulation for angle $\phi_2^{54:53}$ related to Atlas' current orbit (red star in Figure~\ref{fig_1}).} 
	\label{fig_2}%
\end{figure}

The domains of the 54:53 Prometheus-Atlas resonance resemble the shaped of an “hourglass”. Figures~\ref{fig_1}(b) provides a detailed view of the Corotation (\textbf{C}) and Lindblad (\textbf{L}) zones associated with the 54:53 Prometheus-Atlas MMR. The Corotation zone appears as a well-defined whitish “central spot” in the range [138,321 km, 138,326 km] and [0.0035, 0.0065] as detailed in Figure~\ref{fig_1}(c). In the \textbf{C} region, the critical angle $\phi_2  = 54\lambda_{Pro} - 53\lambda_S - \varpi_{Pro}$ (known as the Corotation angle) oscillates around 180°, where $\lambda$ represents the mean longitude, $\varpi$ is the longitude of pericenter, \textit{Pro} and \textit{S} indicate Prometheus and a clone of Atlas, respectively. The map reveals that the interior of the Corotation zone generally appears whitish, with small values of the spectral number \textbf{N}. This occurs when the motion is regular due to resonance trapping (Figure~\ref{fig_2}(b,d)). In the \textbf{L} region the angle $\phi_1  = 54\lambda_{Pro}-53\lambda_S-\varpi_S$ (referred to as the Lindblad angle) librates around 0 for the initial condition within this region (see Figure~\ref{fig_2}(a,c)).

In Figure~\ref{fig_1}(a) the regions corresponding to the 53:52 and 55:54 Prometheus-“Atlas clone” mean-motion resonances are visible. These structures also exhibit an "hourglass" shape, similar to the 54:53 resonance. For the 53:52 resonance, we identified the Lindbald zone \textbf{L1}, but no defined Corotation zone exist. For the 55:54 resonance, \textbf{C2} and \textbf{L2} represent the Corotation and Lindbald zones, respectively. Several higher-order resonances can also observed around Atlas's orbital vicinity, and we identified second-order resonances.

Colored circles in Figures~\ref{fig_1}(a,c) represent the initial conditions of Atlas clones for which the temporal evolution of the angles $\phi_1^{m+1:m}$ and $\phi_2^{m+1:m}$, where $m$ = 52, 53 and 54, are represented in Figure~\ref{fig_2}. The oscillation around 0 for the Lindbald angle is shown in Figures~\ref{fig_2}(a,c), while the oscillation around 180° for the Corotation angle is displayed in Figures~\ref{fig_2}(b,d). The oscillation periods (represented by resonant mode) of each angle are approximately (a) $\sim$1,310.72 days, (b) $\sim$ 1,260.31 days, (c) $\sim$ 1,560.38 days, and (d) $\sim$ 1,560.38 days. Since Atlas's current orbit lies on the border between the Corotation resonance and its separatrix, the temporal variation of the angle $\phi_2^{54:53}$ presents alternation between oscillation and circulation, as seen in Figure~\ref{fig_2}(e).

By analyzing the temporal behavior for the critical angles $\phi_1$ and $\phi_2$ of Atlas clones in the regions \textbf{A}, \textbf{A1}, \textbf{A2} and \textbf{B}, \textbf{B1} and \textbf{B2}, we identified that both angles exhibit circulation.

\subsection{The Individual Dynamical Power Spectra}
\label{3.1}

Quantitative insights into the period distribution in phase space can be obtained by considering the initial semi-major axis of test satellites as a free parameter while keeping the eccentricity, $e_0$, fixed in all numerical simulations. The result of this analysis is the “Individual Dynamical Power Spectra”, or IPS. On the y-axis of an IPS, for each initial condition, we show the periods associated with peaks in the spectrum that have amplitudes greater than a predefined reference amplitude (RA).

\begin{figure}[!ht]
	\centering 
	\includegraphics[width=1.0 \columnwidth,angle=0, scale = 1.0]{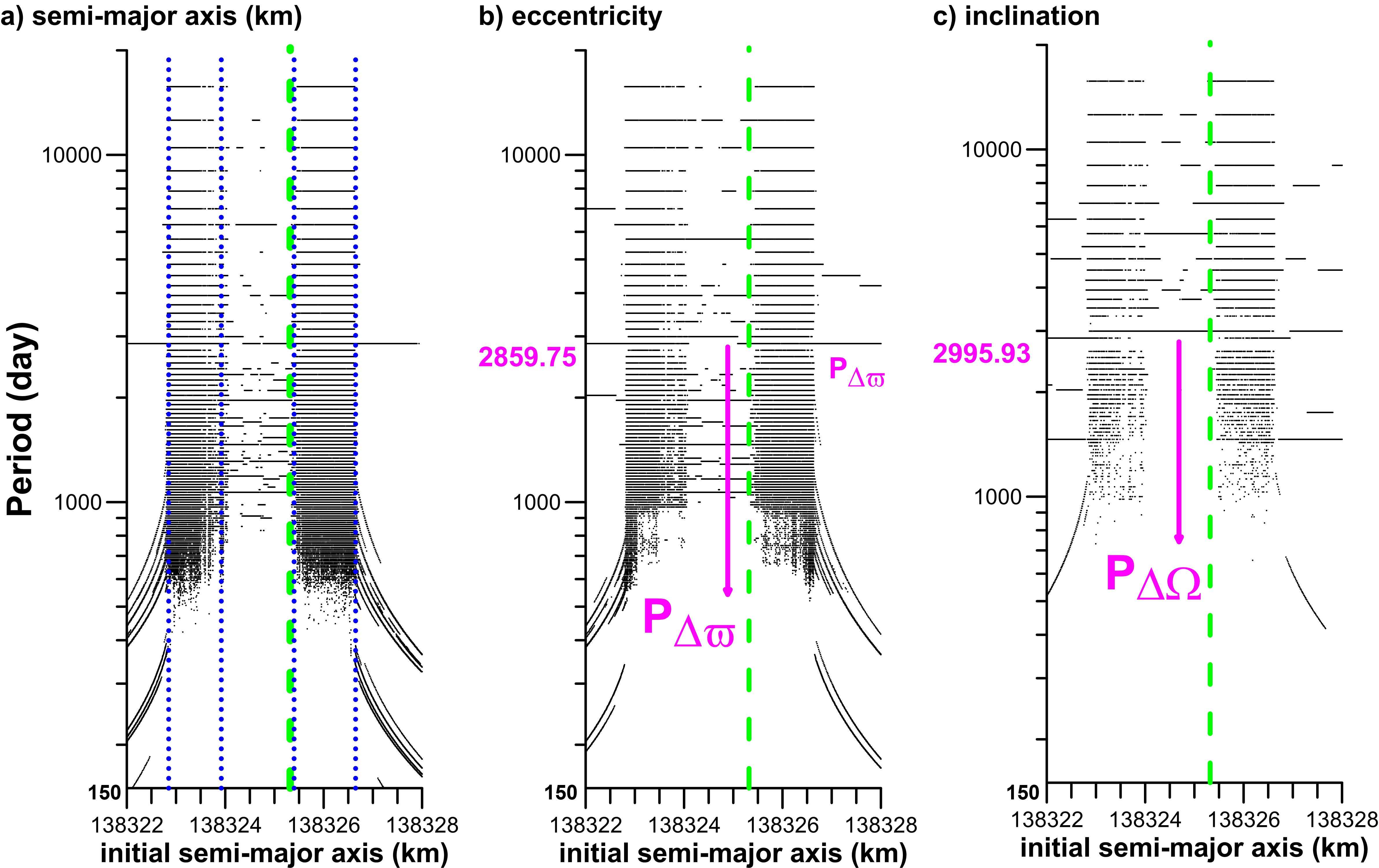}	
	\caption{Individual Dynamical Power Spectra of 6,000 Atlas clones, considering Prometheus as disturbing, constructed from the spectrum of different osculating variables (a) semi-major axis, (b) eccentricity, and (c) orbital inclination. The green dotted line represents the current semi-major axis of Atlas on January 01, 2000. In (a) the blue dotted lines delimit the borders between the Corotation zone (\textbf{C}) and the separatrices, the dark blue contour around \textbf{C} in Figure~\ref{fig_1}. The integration time for each orbit is 172.25 years, with a time sampling of 0.06 days, and only periods greater than 150 days were considered.} 
	\label{fig_3}%
\end{figure}

Figure~\ref{fig_3} shows three IPS results, each obtained from the spectra calculated for the osculating variables: (a) semi-major axis, (b) eccentricity, and (c) orbital inclination of 6,000 Atlas clones within the interval 138,322 km $\leq$ $a_0$ $\leq$ 138,328 km, RA is set to 5\%, and the y-axis is presented on a logarithmic scale. The initial osculating eccentricity of Atlas on January 1st, 2000, indicated by the red line $e_0$ = 0.0059 in Figure~\ref{fig_1}(c) was fixed. The loci of the fundamental periods for eccentricity and inclination are distributed vertically and identified with the respective symbols ($P_{\Delta\varpi}$, $P_{\Delta\Omega}$ as discussed in Section~\ref{3.2}).

In Figure~\ref{fig_3}(a), blue dotted lines mark regions with a high number of peaks associated with irregular motion, corresponding to the separatrices of the 54:53 resonance shown in Figure~\ref{fig_1}(c). The green dotted line in Figure~\ref{fig_3} represents the current semi-major axis of Atlas, which is situated on the border of the Corotation zone.

\begin{figure}[!ht]
	\centering 
	\includegraphics[width=1.0 \columnwidth,angle=0, scale = 0.90]{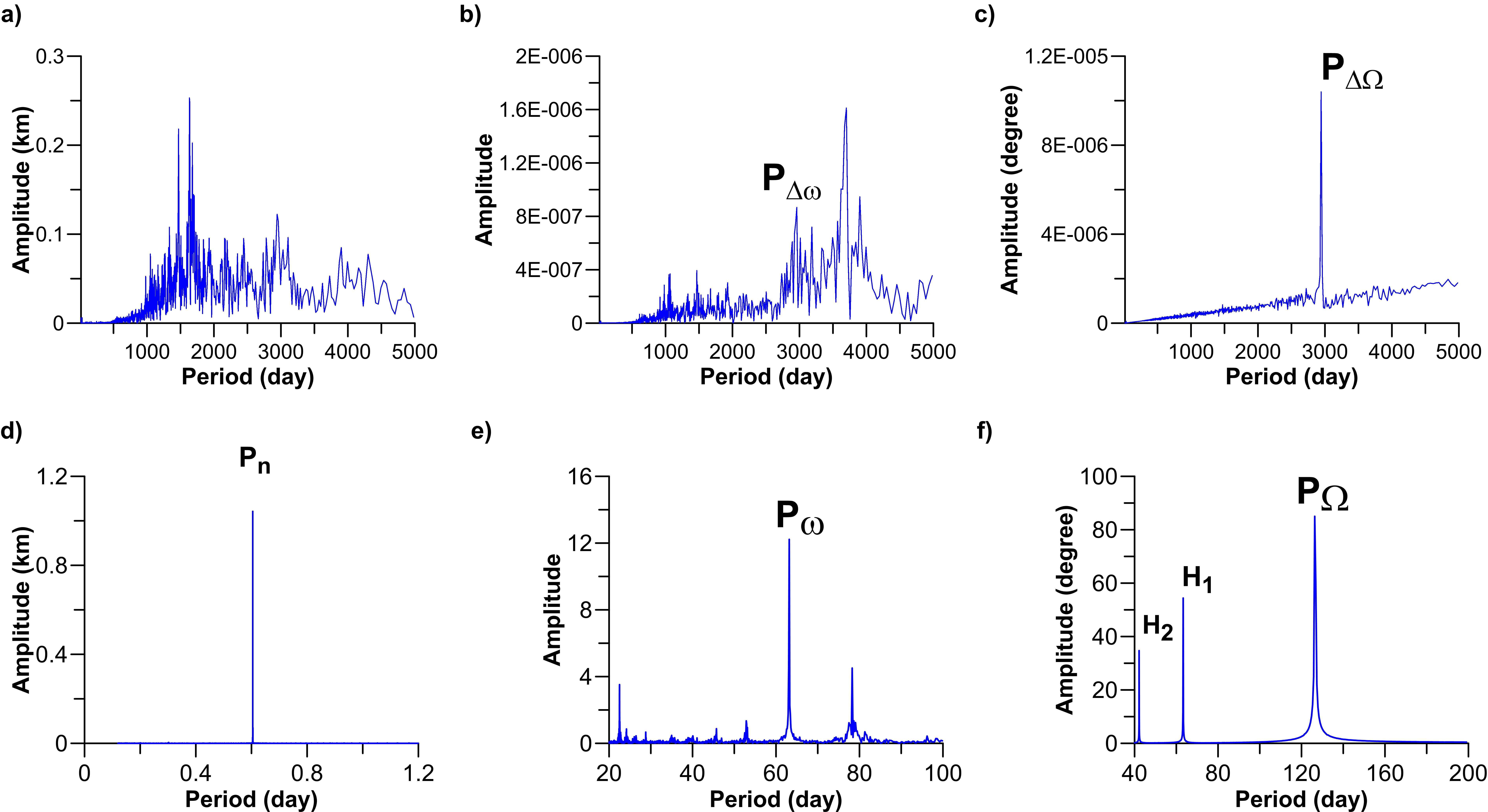}	
	\caption{Individual spectra for the current osculating orbit of Atlas on January 1st, 2000. (a,d) semi-major axis, (b) eccentricity, (c) orbital inclination, (e) argument of pericenter, and (f) ascending node. These spectra were obtained as in Figure~\ref{fig_3}. The high number of peaks observed in (a-c) is related to the current orbital position of Atlas. In (b) and (c) the respective secular period in the eccentricity ($P_{\Delta\varpi}$$\sim$ 2.859,75 days) and in the inclination ($P_{\Delta\Omega}$$\sim$ 2.995,93 days) can be noted. In (d) the orbital period ($P_{n}$$\sim$ 0.6 days), (e) the period for the argument of the pericenter circulation ($P_\omega$ $\sim$ 63 days) and (f) the period for the ascending node ($P_\Omega$ $\sim$ 130 days) and the respective harmonics $H_1$ = $\frac{1}{3}$$P_\Omega$, $H_2$ = $\frac{1}{2}P_\Omega$ and $H_3$ = $\frac{1}{2}P_\omega$ can be identified.} 
	\label{fig_4}%
\end{figure}

\begin{figure}[!ht]
	\centering 
	\includegraphics[width=1.0 \columnwidth,angle=0, scale = 0.68]{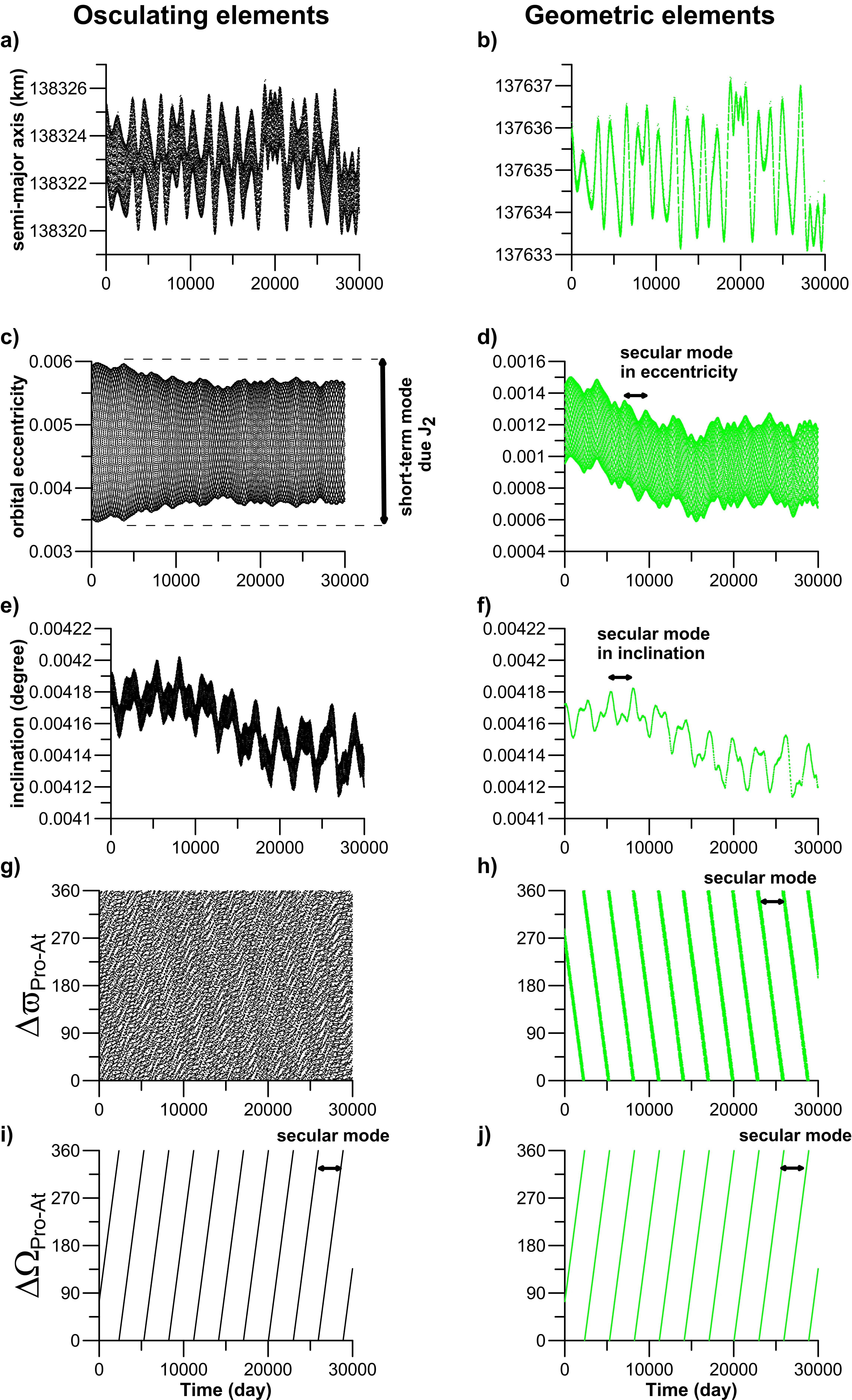}	
	\caption{Comparison between the osculating elements (on the left) and geometric elements (on the right) for the temporal variation of (a,b) semi-major axis, (c,d) eccentricity and (e,f) orbital inclination (degree) relative to Saturn's equator. In (c) we see the amplitude of $\sim$0.0025, the effect of the short-term component caused by $J_2$ on the osculating eccentricity. We note how the rapid circulation of the pericenter of Atlas affects the (g) osculating $\Delta\varpi_{Pro-At}$ making its interpretation difficult. (h) geometric $\Delta$$\varpi_{Pro-At}$, (i) osculating and (j) geometric $\Delta$$\Omega_{Pro-At}$. We observe the long-period oscillations ($\sim$ 3 thousand days) for the geometric eccentricity and inclination caused by the secular component. This simulation used the starting date January 1st, 2000, sampled every 1 day. Satellites included Atlas and Prometheus, along with terms $J_2$, $J_4$, and $J_6$.} 
	\label{fig_5}%
\end{figure}

When the spectrum contains a high number of peaks, the orbital motion can be considered quasi-periodic, irregular, or chaotic. In such cases, the amplitudes are often not evenly spaced, making it challenging to identify the principal periods. An example of this behavior is illustrated in the upper part of Figure~\ref{fig_4} which shows the spectrum for the osculating variables: (a) semi-major axis, (b) the eccentricity, and (c) the orbital inclination of Atlas, with initial conditions set to January, 1st, 2000. In Figures~\ref{fig_4}(b,c) the periods $P_{\Delta\varpi}$ and $P_{\Delta\Omega}$ corresponding to the circulation of the angles $\Delta\varpi\equiv\varpi_{Pro}-\varpi_{At}$ and $\Delta\Omega\equiv\Omega_{Pro}-\Omega_{At}$, respectively, are observed (see Figures~\ref{fig_5}(h,j)). The left and right columns in Figure~\ref{fig_5} show, respectively, the time variations of osculating and corresponding geometric elements: (a,b) semi-major axis, (c,d) eccentricity, (e,f) inclination concerning Saturn’s equator, (g,h) relative longitude of pericenters $\Delta\varpi_{Pro-At}$ and (i,j) relative longitude of ascending nodes $\Delta\Omega_{Pro-At}$.
The periods of $\Delta\varpi$ and $\Delta\Omega$ are centered around $\sim$3.000 days and will be analyzed further in Section ~\ref{3.2}.

Figure~\ref{fig_4}(d) displays a portion of the short-period spectrum for the semi-major axis, where the orbital period ($P_n$ $\sim$ 0.6 day) is identified. In Figures~\ref{fig_4}(e,f), the spectra for the argument of the pericenter and the ascending node are shown, revealing the periods associated with the circulation of the pericenter ($P_\omega$$\sim$ 63 days) and the circulation of the ascending node ($P_\Omega$ $\sim$ 130 days), along with respective harmonics $H_1$ = $\frac{1}{2}$$P_\Omega$, $H_2$ = $\frac{1}{3}$$P_\Omega$ and $H_3$ = $\frac{1}{2}$$P_\omega$. The spectral analysis of these elements indicates that the highest amplitudes are linked to short-period frequencies.

Next, we will analyze the long-period variations in Atlas's orbit caused by the secular relationship with Prometheus and the variations due to $J_2$ (see Figure~\ref{fig_5}(c) for an example).

\subsection{Secular and $J_2$ short-term variations in Atlas's current orbit}
\label{3.2}

The eccentricity and inclination variables have two key components, which are related to the following angles: $\Delta\varpi\equiv\varpi_{Pro}-\varpi_{At}$ and $\Delta\Omega\equiv\Omega_{Pro}-\Omega_{At}$, respectively (Brouwer and Clemence 1966). Callegari et al. (2021) define the fundamental frequencies associated with these angles as the mutual-secular long-term mode (in this study, this component is referred to as the secular mode). These frequencies are linked to the secular relative rates of the pericenters and ascending nodes due to mutual perturbations, as well as Satrun's $J_2$, $J_4$, $J_6$ terms for both Atlas and Prometheus (for further details, see Callegari et al. 2021). In the case of the Prometheus-Atlas system, the associated periods are $P_{\Delta\varpi}$ $\sim$ 2.859,75 days and $P_{\Delta\Omega}$ $\sim$ 2.995,93 days. The plots of eccentricity, inclination, $\Delta\varpi_{Pro-At}$ and $\Delta\Omega_{Pro-At}$ are shown in Figure~\ref{fig_5}.

The loci for the periods $P_{\Delta\varpi}$ and $P_{\Delta\Omega}$ are displayed in the spectra of the eccentricity and inclination in Figures~\ref{fig_3}(b,c). It is important to note that $P_{\Delta\varpi}$ and $P_{\Delta\Omega}$ are interrupted at the borders of the resonance but reappear beyond them, as they are not independent of resonance. In Figure~\ref{fig_5}(g) the osculating argument $\Delta\varpi_{Pro-At}$ appears as a stripe, where the temporal variation for this argument is overshadowed by the short-term $J_2$ component of the pericenter of both satellites. Conversely, Figure~\ref{fig_5}(h) shows that the geometric argument $\Delta\varpi_{Pro-At}$ circulates in a retrograde sense with period $P_{\Delta\varpi}$. Figures~\ref{fig_5}(i,j) reveal that both the osculating and geometric arguments $\Delta\Omega_{Pro-At}$ circulates in a prograde sense with period $P_{\Delta\Omega}$, as $\Omega$ is not affected by short-term $J_2$ variations.

\begin{figure}[!ht]
	\centering 
	\includegraphics[width=1.0 \columnwidth,angle=0, scale = 0.85]{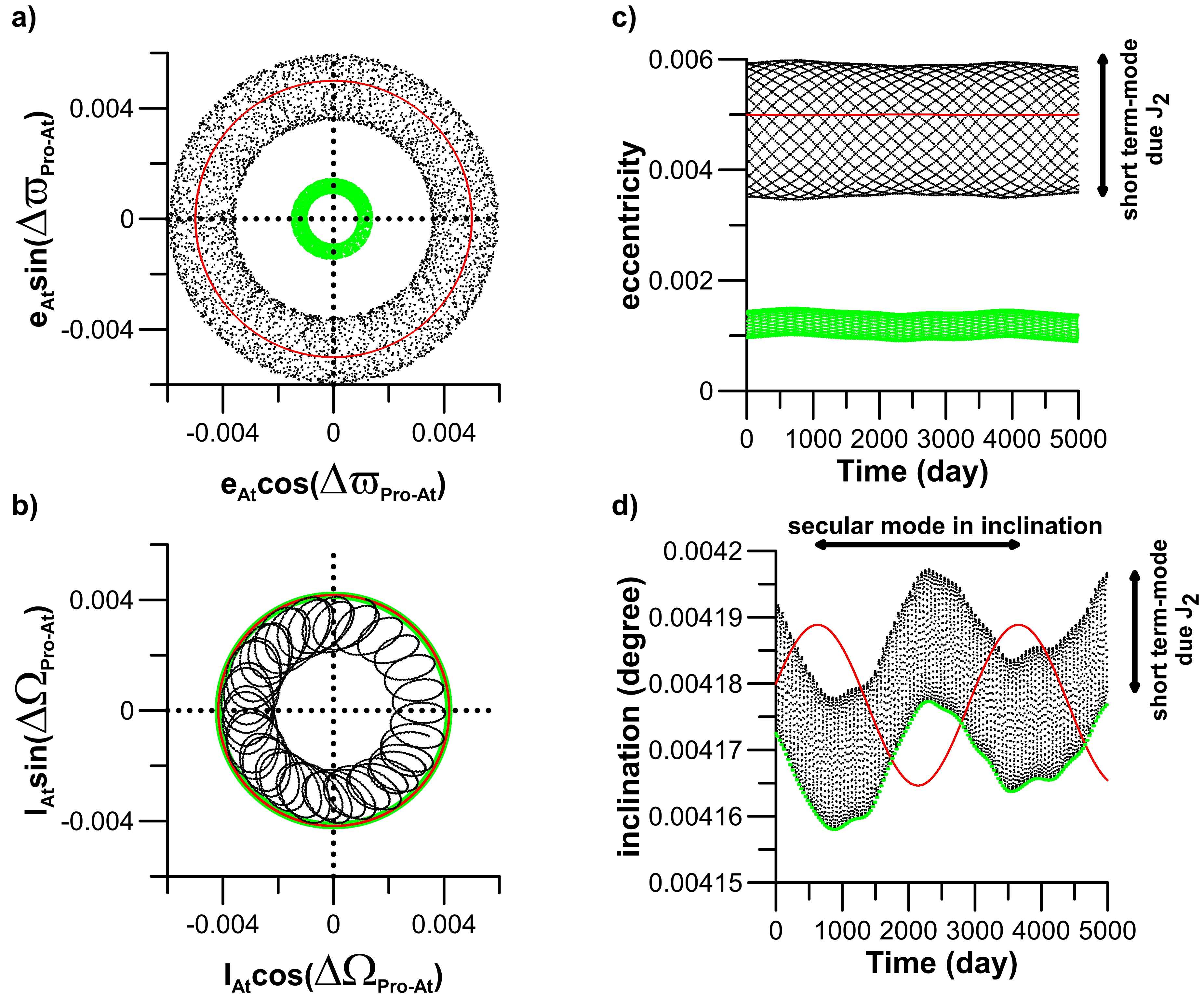}	
	\caption{\textbf{Osculating (black curve), secular (red curve), and geometric (green curve) eccentricity and inclination for Atlas.} The red curves were obtained with classical secular theory including Saturn, Prometheus, and the coefficients $J_2$ and $J_4$. In (a) and (b) the polar representation and the absence of the forced term for these variables can be seen. The effects caused by the term $J_2$ on these variables can be identified in (c) and (d).} 
	\label{fig_6}%
\end{figure}

By comparing Figures~\ref{fig_5}(a,b), we observe that the osculating semi-major axis is influenced by $J_2$ perturbation, resulting in a larger semi-major axis compared to the geometric semi-major axis. The same effect is visible in the eccentricity (see Figures~\ref{fig_5}(c,d)), and inclination, (see Figures~\ref{fig_5}(d,e)). On January 1st, 2000, the initial values of these quantities were approximately $a_0$$\sim$ 138.325,32 km and $a_g$$\sim$ 137.635,81 km ($\Delta$$a$$\sim$ 698,51 km); $e_0$$\sim$ 0.0059 e $e_g$$\sim$ 0.001 ($\Delta$$e$$\sim$ 0.005); $i_0$$\sim$ 0.00419 e $i_g$$\sim$ 0.00418 ($\Delta$$i$$\sim$ 0.00001), where $o$ and $g$ refer to osculating and geometric elements, respectively, and $\Delta$ represents the numerical difference between these quantities. These osculating elements exhibit short-term variations induced by the $J_2$ mode, producing amplitudes in the respective temporal variations (see Figures~\ref{fig_6}(c,d)).

The red curves in Figure~\ref{fig_6} represent the analytical solutions of secular theory (Murray and Dermott 1999, Chapter 7) for the eccentricity and inclination of Atlas. The black, green, and red curves in Figures~\ref{fig_6}(a,b) showed the projection of the osculating, geometric, and secular orbits of Atlas onto the planes ($e_{At}$$\cos⁡(\Delta\varpi_{Pro-At})$, $e_{At}$$\sin⁡(\Delta\varpi_{Pro-At})$) and ($I_{At}$$\cos⁡(\Delta\Omega_{Pro-At})$, $I_{At}$$\sin⁡(\Delta\Omega_{Pro-At})$). From Figures~\ref{fig_6}(a,b) we can be seen that the forced terms for eccentricity and inclination are negligible.

Figures~\ref{fig_6}(c,d) provide detailed views of the eccentricity and inclination depicted in Figure~\ref{fig_5}. The $J_2$ component causes a variation of $\sim$ 0.0025 in the osculating eccentricity and $\sim$ 0.00002 degrees in the osculating inclination, as seen in Figures~\ref{fig_5}(c,e). As predicted, the geometric inclination remains unaffected by this perturbation.

\section{Conservative dynamics II: The influence of Prometheus's eccentricity on Atlas's orbit and the 70:67 Pandora-Atlas resonance}
\label{4}

\subsection{The influence of Prometheus's eccentricity on the 54:53 mean-motion resonance}
\label{4.1}

As described in Section~\ref{3}, Atlas's orbit lies on the Corotation border zone with the separatrices related to the 54:53 resonance. Renner et al. (2016) suggest that Prometheus's orbital eccentricity is a key parameter for the stability of the 54:53 MMR. If Prometheus's geometric eccentricity is less than or equal to 0.00008, Atlas's orbit would be trapped within the resonance.

\begin{figure}[!ht]
	\centering 
	\includegraphics[width=1.0 \columnwidth,angle=0, scale = 0.9]{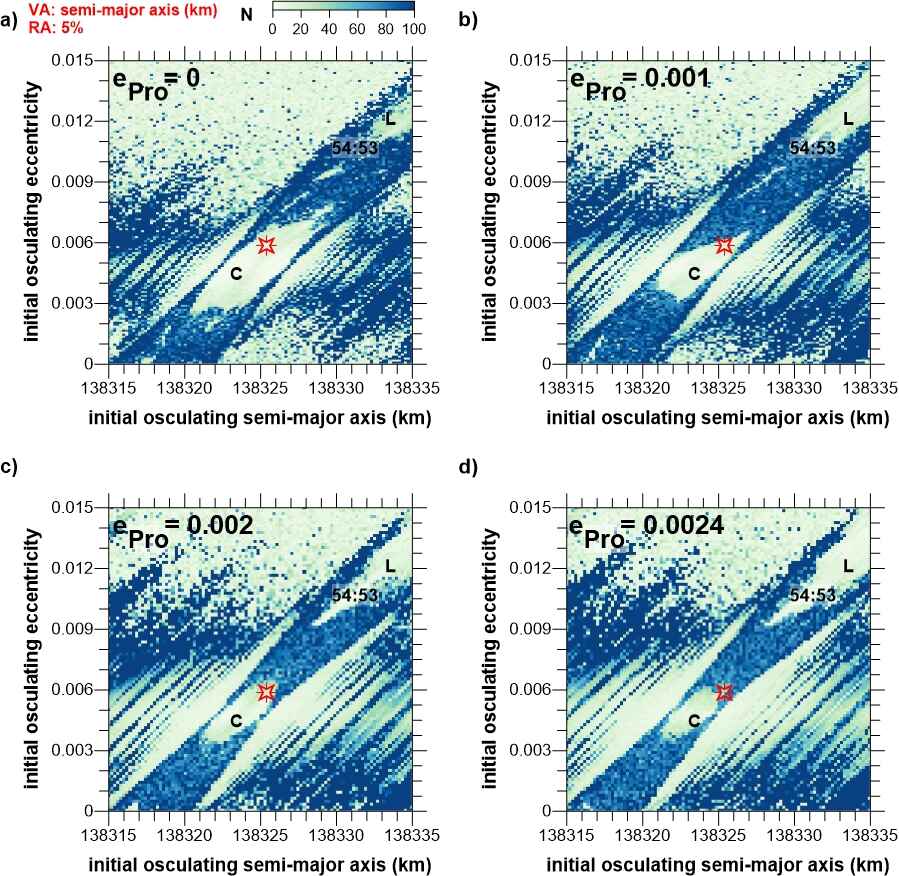}	
	\caption{Similar to Figure~\ref{fig_1}(b), therefore the orbits of Atlas clones had been obtained by integrating the Prometheus clone with osculating eccentricity equal to (a) 0, (b) 0.001, (c) 0.002, and, (d) 0.0024. \textbf{C} and \textbf{L} represent the Corotation and Lindbald zones associated with the 54:53 "Prometheus clone"-Atlas resonance. The current Atlas's orbit (red star) is inside zone \textbf{C} in (a) and (b). In (d) Atlas is closer to the Corotation boundary.} 
	\label{fig_7}%
\end{figure}

Figure~\ref{fig_7} illustrates the domain of the 54:53 MMR by considering a Prometheus clone with varying osculating eccentricity: (a) 0, (b) 0.001, (c) 0.002, and (d) 0.0024.

When comparing Figure~\ref{fig_1}(b) with Figures~\ref{fig_7}(a,b), it becomes apparent that Atlas lies within the Corotation region due to the small eccentricities of the Prometheus clone. The Corotation zone is larger than what is seen in Figure~\ref{fig_1}(b). However, in Figures~\ref{fig_7}(c,d) representing a Prometheus clone with an eccentricity close to its current value, the Corotation zone contracts and is filled by the separatrices, placing Atlas near the border between the Corotation zone and its separatrices. At the same time, the Lindblad zone expands as the eccentricity of the Prometheus clone increases.

\subsection{The 70:67 Pandora-Atlas resonance}
\label{4.2}

In addition to this proximity to the 54:53 Prometheus-Atlas resonance, Atlas's current orbit experiences minor perturbation due to Pandora. Spitale et al. (2006) proposed the existence of a 70:67 Pandora-Atlas commensurability to explain these perturbations. In this section, we explore this perturbation in Atlas's orbit using the same tools applied previously.

\begin{figure}[!ht]
	\centering 
	\includegraphics[width=1.0 \columnwidth,angle=0, scale = 0.85]{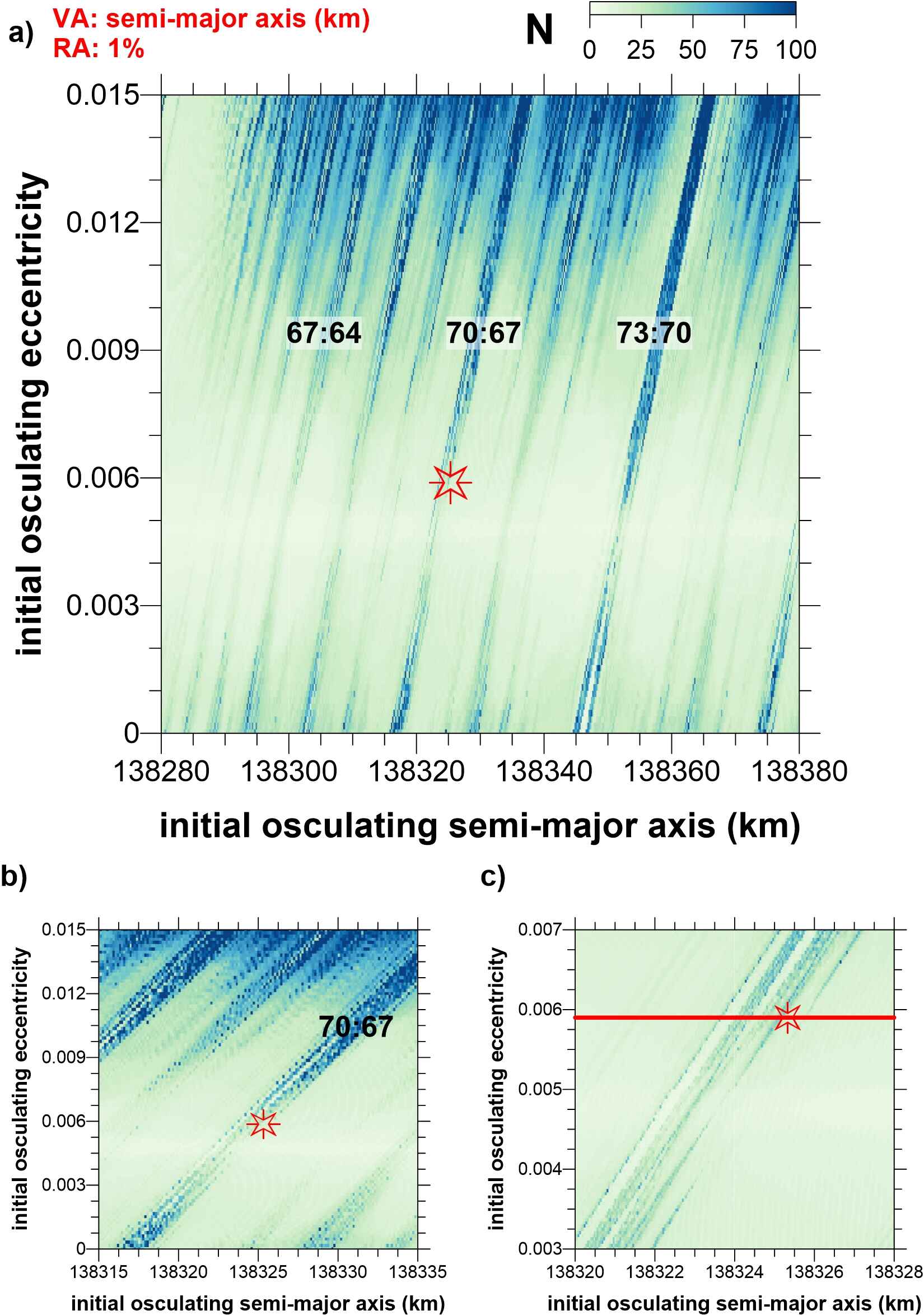}	
	\caption{(a) The same as Figure~\ref{fig_1}, where Pandora replaces Prometheus as Atlas disturbing body. The locations of several resonances near the Atlas (red star) can be seen. (b) and (c) detail around the Atlas current position and the 70:67 resonance.} 
	\label{fig_8}%
\end{figure}

\begin{figure}[!h]
	\centering 
	\includegraphics[width=1.0 \columnwidth,angle=0, scale = 0.75]{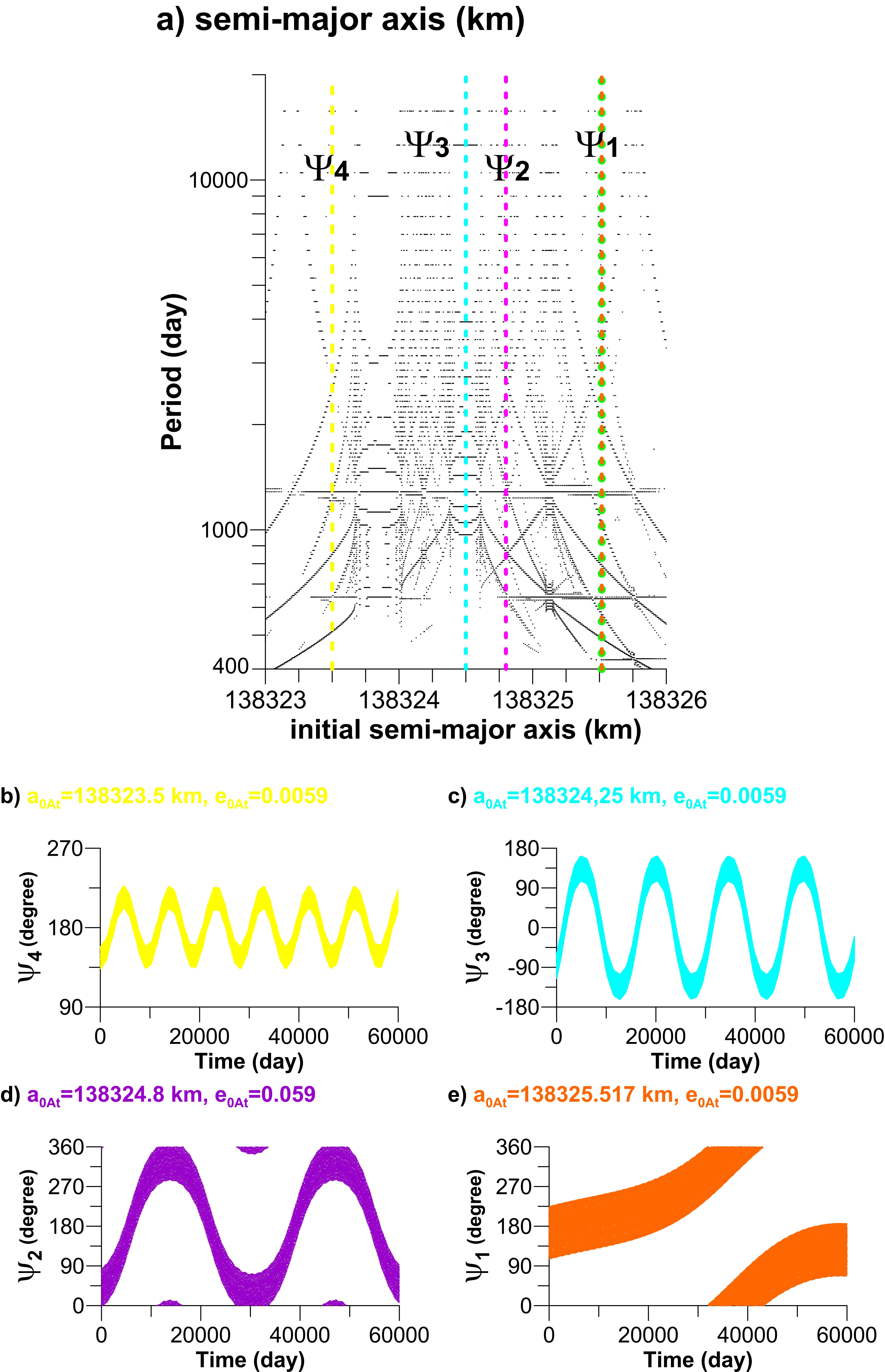}	
	\caption{(a) Same as Figure~\ref{fig_3}(a) considering Pandora as the only disturbing satellite and RA=0.1\%.The domains for the multiplets $\Psi_{4}$, $\Psi_{3}$, $\Psi_{2}$ and $\Psi_{1}$ can be identified. Temporal variation for the geometric arguments (b) $\psi_{4}$, (c) $\psi_{3}$, (d) $\psi_{2}$ and (e) $\psi_{1}$ and the colored vertical lines in Figure ~\ref{fig_9}(a) represent the respective values for the semi-major axis of these Atlas clones.} 
	\label{fig_9}%
\end{figure}

Figure~\ref{fig_8} and Figure~\ref{fig_9} were constructed similarly to Figure~\ref{fig_1} and Figure~\ref{fig_3}(a), respectively, but with Pandora as the perturber instead of Prometheus. The 70:67 resonance appears as a diagonal band in Figure~\ref{fig_9}, containing a narrow central with a small value for spectral number \textbf{N} ($0\leq$\textbf{N}$\leq20$), surrounded by the separatrices (see Figure~\ref{fig_8}(a)). Atlas's current orbit is represented by the red star, positioned at the boundary of this resonance's separatrix, as shown in Figure~\ref{fig_9}(a) (green dashed vertical line). Additionally, two other resonances involving Pandora and Atlas clones, namely 67:64 and 73:70, are visible in the phase space, located further from Atlas's current orbit (the red star in Figure~\ref{fig_8})

Figure~\ref{fig_9}(a) shows the loci for the $\Psi_{4}$, $\Psi_{3}$, $\Psi_{2}$ and $\Psi_{1}$ multiplets associated with the 70:67 resonance.The overlap of some multiples (described by Renner et al. 2016) forms vertical barriers with numerous peaks, indicating the irregular nature of these orbits. This irregularity arises because these phase space regions are located within the resonance separatrices (see Figure~\ref{fig_8}(c)). The green dashed vertical line in Figure~\ref{fig_9}(a) represents the current semi-major axis of Atlas, which is located near the boundary of the $\Psi_{1}$ multiplet.

Figure ~\ref{fig_9} depicts the temporal variation of the geometric arguments: (b) $\psi_{4}=70\lambda_{Pa}-67\lambda_{S}-3\varpi_{Pa}$, (c) $\psi_{3}=70\lambda_{Pa}-67\lambda_{S}-2\varpi_{Pa}-\varpi_{S}$, (d) $\psi_{2}=70\lambda_{Pa}-67\lambda_{S}-2\varpi_{S}-\varpi_{Pa}$, (e) $\psi_{1}=70\lambda_{Pa}-67\lambda_{S}-3\varpi_ {S}$, where \textit{Pa} refers to Pandora and \textit{S} to  Atlas clone with initial condition ($a_{0}$, $e_{0}$) observed in Figures~\ref{fig_9}(b-e). The oscillation of $\psi_{4}$ and $\psi_{3}$ around 180° and 0° respectively, are noted, while $\psi_{2}$ and $\psi_{1}$ are observed circulating. 

\begin{figure}[!ht]
	\centering 
	\includegraphics[width=1.0 \columnwidth,angle=0, scale = 0.9]{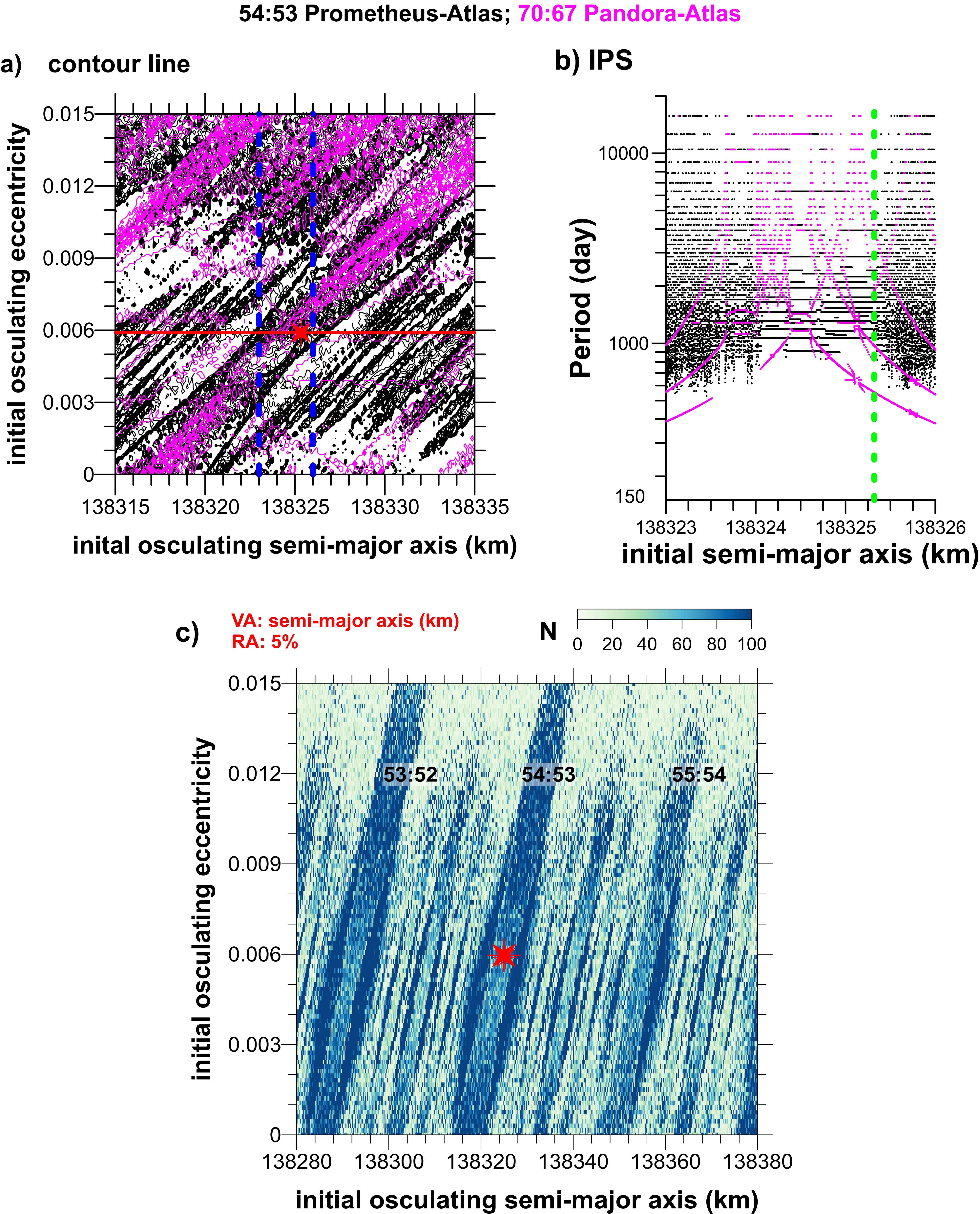}	
	\caption{(a) Contour lines of the spectral number \textbf{N} for subsystems including only Prometheus (black) or Pandora (magenta) as disturbers of Atlas. (b) IPS for  Atlas's current eccentricity (indicated by the red vertical line in (a)), bounded in the region [138,323 km, 138,326 km] (blue vertical lines in (a)). (c) The same as Figure~\ref{fig_1}(a), including Pandora perturbation.} 
	\label{fig_10}%
\end{figure}

Inspection of Figure~\ref{fig_1} and Figure~\ref{fig_8} suggests that frequencies related to the 54:53 Prometheus-Atlas and 70:67 Pandora-Atlas resonances overlap in the same phase space. To visualize how the overlap affects Atlas's orbit, we determined the contour maps of the spectral number \textbf{N}. In Figure~\ref{fig_10}(a), the black (magenta) contour lines represent the interactions between Prometheus-Atlas (Pandora-Atlas). The “hourglass” structure is visible and the Corotation and Lindblad regions associated with the 54:53 resonance are visible. The magenta contour lines penetrate the Lindblad zone and fill the Corotation zone, indicating the overlap between the two resonances. A similar result is seen in the IPS shown in Figure~\ref{fig_10}(b), which combines the IPS of Figure~\ref{fig_3}(a) and Figure~\ref{fig_9}(a).

Figure~\ref{fig_1} and Figure~\ref{fig_8} enable us to identify the key areas in phase space associated with the resonant perturbations caused by Prometheus and Pandora, respectively. Thus, it is natural to consider a mapping that incorporates the combined perturbative effects of Prometheus and Pandora. Figure~\ref{fig_10}(c), constructed similarly to Figure~\ref{fig_1}(a), represents this dynamic scenario. When comparing Figure~\ref{fig_10}(c) and Figure~\ref{fig_1}(a), it becomes clear that the regular domains related to the Corotation and Lindblad resonances, which were sharply defined in the individual maps (Figure~\ref{fig_1} and Figure~\ref{fig_8}), have been replaced by a diffuse structure in the phase space typical of chaotic systems.

In conclusion, the overlapping perturbations from 54:53 Prometheus-Atlas and 70:67 Pandora-Atlas resonances in the phase space, amplify the perturbations in Atlas's orbital dynamics (for more detail see ~\ref{A2} and Figure~\ref{fig_A1}). In the next section, we will quantify the chaos in Atlas's orbit with greater precision.

\section{The chaotic orbit of Atlas }
\label{5}

\subsection{Chaotic diffusion}
\label{5.1}

In phase space regions, where complex phenomena arise, such as near the border between a strong resonance and its stochastic layer (separatrix), a detailed stability analysis is necessary. To quantify the chaotic diffusion within a given region and assess the long-term stability due to slow chaotic diffusion (or weak chaos), we approach the problem of Atlas's dynamical stability by mapping the chaotic diffusion coefficient along a section of the system's phase space with various configurations. This approach us to better understand the dynamical characteristics of finer regions within the phase space.

\subsubsection{Chaotic Diffusion Coefficient}
\label{5.1.1}

Chaotic diffusion in non-linear systems is commonly studied in the action domain. The sizes of chaotic regions in the phase space can be accurately estimated through a frequency-domain study (Laskar 1990). One advantage of using frequency space is the relationship between the diffusion of independent frequencies in the action space and the diffusion of independent frequencies in time. Laskar (1993) showed that this relation takes the form of the Diffusion Equation
\begin{equation}\label{eq1}
    \nabla^2f(x,t)\propto\frac{\partial}{\partial{t}}f(x,t),
\end{equation}
hereafter written as $\delta\delta_x\propto\delta_tf(x,t)$.

This equation provides two diffusion coefficients obtained from the frequency-space approach: one related to the diffusion in space (left-hand side of the equation) and the other in time (right-hand side of the equation).

The diffusion of frequencies in space can be determined by calculating the second derivative of the main amplitude frequency in the IPS, which makes it sensitive to the numerical parameters used to construct the IPS, such as total integration time (T) and sampling rate ($\Delta$t).

On the other hand, the diffusion of frequencies in time has been studied more extensively. In Laskar (1993) and Ferraz-Mello and Wachlin (1998), the diffusion coefficient in time was defined as $\delta_tf(x,t)$ = $|\nu^{2}-\nu^{1}|$ where $\nu^{1}$ is the frequency of highest amplitude obtained from the Fast Fourier Transform (FFT) applied to the data in the first half of the total integration time and $\nu^{2}$ is the highest amplitude frequency for the second half of the total integration time.

Since the FFT provides information in the frequency domain but not in time, we applied the Wavelet Analysis Method (WAM) as described by Michtchenko and Nesvorný (1996) The WAM proved sensitive to both to weak and strong chaotic motion, allowing us to track the time evolution of independent frequencies.

After determining the independent frequency in both the frequency and time domain, we redefined the diffusion coefficient in space as $\delta_tf(x,t)$ = $|\nu^{{\prime}{2}}-\nu^{{\prime}{1}}|$, where $\nu^{{\prime}{1}}$ and $\nu^{{\prime}{2}}$ are analogous to the original definition, but with the mean frequencies calculated from the application of the WAM to the first and second halves of the total integration timespan (Guimarães and Michtchenko 2023).

\subsubsection{Methodology}
\label{5.1.2}

Given the numerical limitations shared between the diffusion coefficient in space and IPS, we focus only on the diffusion coefficient of frequencies in time. Therefore, $\delta_tf$ will be applied instead of $\delta_tf(x,t)$.

Once the diffusion coefficients are calculated, they will be interpreted based on the methodology of Marzari et al. (2003): i) regular motion regions are characterized by $\delta_tf<10^{-5}$, ii) regions with $\delta_tf>10^{-1}$ are regarded as highly chaotic (fast chaotic diffusion), and iii) regions between $10^{-5}<\delta_tf<10^{-1}$ are considered weakly chaotic (slow chaotic diffusion) or quasi-periodic and can be considered long-term stable.

The coefficient $\delta_tf$ is calculated for a set of initial conditions in the semi-major axis 138,323 km $<$ $a_0$ $<$ 138,327 km corresponding to the extension of the Corotation resonance (Figure~\ref{fig_1}), with a fixed initial eccentricity $e_0$=0.0059. Atlas clones were integrated for 100 years for all the initial conditions. The coefficients were calculated using the temporal evolution of the frequency related to the semi-major axis of Atlas $f_a$ and the eccentricity $f_e$ (black and red curves in Figure~\ref{fig_11}, respectively). These were calculated for three distinct dynamical systems: (a) Atlas and Prometheus, (b) Atlas and Pandora, and (c) Atlas, Prometheus, and Pandora (top, middle and bottom plot of Figure~\ref{fig_11}, respectively).

\subsection{Atlas-Prometheus}
\label{5.2}

Figure~\ref{fig_11}(a) illustrates the behavior of the diffusion coefficient along the semi-major axis span of the Corotation resonance.

Analysis of the coefficient shows that Atlas's dynamics are governed by slow diffusion across most of the region. Near Atlas's nominal position (vertical green dotted line in $a_0$=138,325.32 km) the diffusion coefficient is $\delta_tf\approx10^{-3}$, consistent with slow chaotic diffusion in the border region between the resonance and the stochastic layer, where complex phenomena occur.

By comparing the diffusion coefficient along the section with the results shown in the IPS in Figure~\ref{fig_3}, it become evident that, although the motion is characterized as chaotic due to the high number of frequency peaks detected by the chaos detection tool, the motion in the region is dominated by the system's secular evolution. The higher-frequency (short-period) terms are easily identifiable, and the chaotic region is confined to the low-frequency region (long-period terms, $P>$ 1,000 days).

This confined spreading in the frequency space, combined with the calculated diffusion coefficient, supports the arguments for long-term stability.The complex structure of the phase space suggests that Atlas is at the border of the resonance and diffuses slowly across the stochastic layer between the resonances, potentially becoming trapped there. 

\subsection{Atlas-Pandora}
\label{5.3}

In the Atlas-Pandora system (without Prometheus), the phase space of Atlas is strongly modified, as seen in Figure~\ref{fig_8}, leading to a change in the behavior of the diffusion coefficients.(Figure~\ref{fig_11}(b)). Since Prometheus is the main perturber of Atlas's orbit, its absence makes Atlas's orbit more dynamically stable.

Indeed, the diffusion coefficient in this scenario is consistently an order of magnitude lower than in the Prometheus-Atlas case, frequently reaching values $\delta_t f<10^{-4}$.

This result aligns with the IPS shown in Figure~\ref{fig_9}(a), where low-frequency (long-period, $P>$ 1000 days) terms are identifiable and high-frequency (short-period, $P<$100 days) terms show a narrow, confined spread around the independent frequencies.

In this case, as demonstrated in Figure~\ref{fig_9} and Figure~\ref{fig_10} that Atlas's evolution is primarily driven by quasi-periodic motion due to the 70:67 resonance with Pandora, indicating that Atlas could reamin in its current configuration for extended periods.

These results are consistent with Cooper et al. (2015), who observed that the Fast Lyapunov Indicators for Atlas’s motions under Pandora's influence grow steadily with a low angular coefficient, indicating lower Lyapunov Exponents compared to the scenario involving both Prometheus and Atlas.

\subsection{Atlas-Prometheus-Pandora}
\label{5.4}

\begin{figure}[!ht]
	\centering 
	\includegraphics[width=1.0 \columnwidth,angle=0, scale = 0.65]{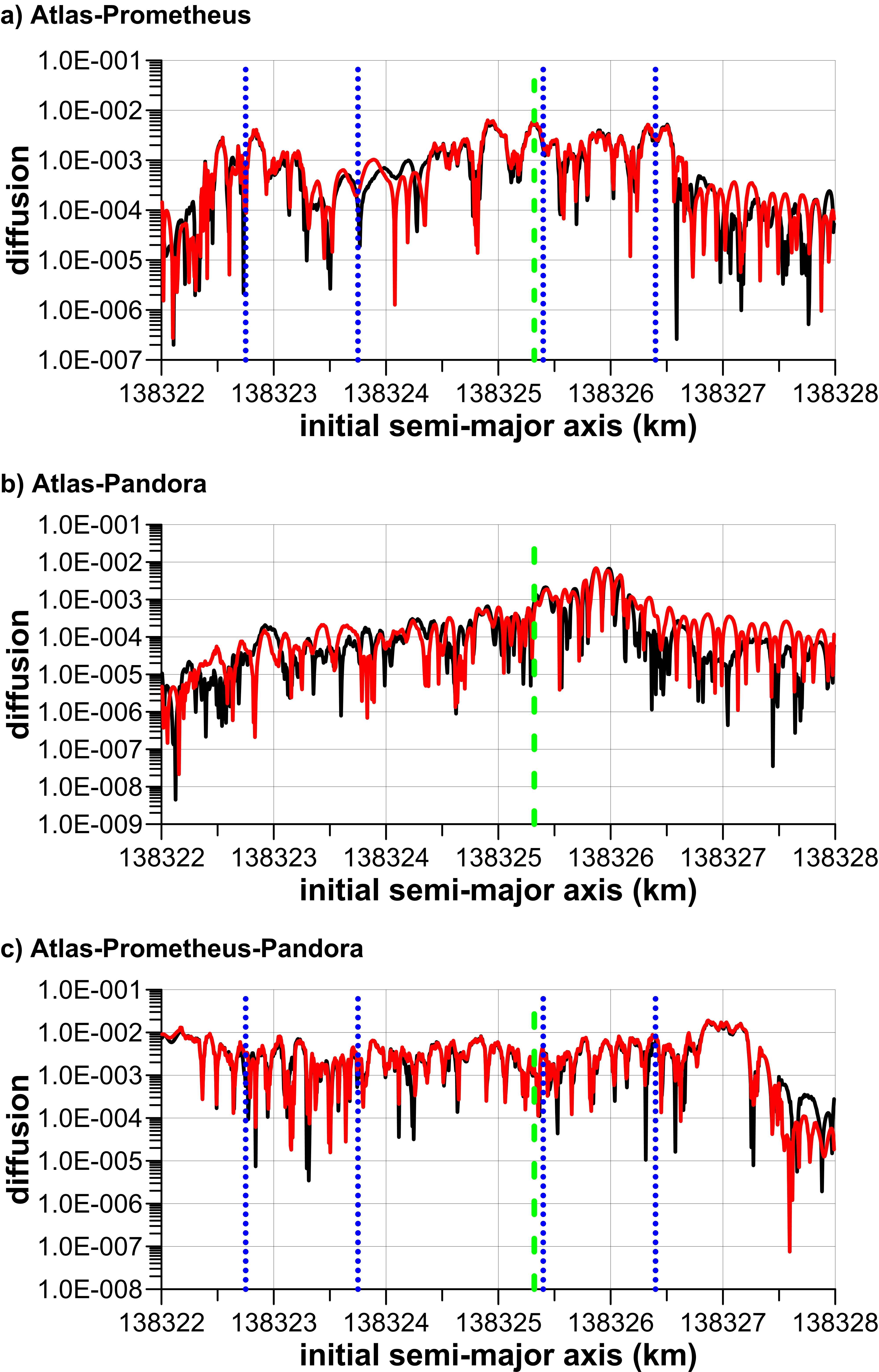}	
	\caption{The blue dotted vertical lines represent the borders between the Corrotation zone and the 54:53 resonance separatrices and, the green dashed vertical line represents the value for the current Atlas semi-major axis. (a) The low value for the diffusion coefficient for the Atlas-Prometheus system around the Atlas vicinity indicates a slow diffusion in its orbital dynamics, revealing that there are no major perturbations that alter the orbital evolution to the semi-major axis (black line) and orbital eccentricity (red line). (b) The diffusion coefficients have been well-behaved in most sections, with a consistently low value, except for the region related to the separatrix of the 70:67 resonance with Pandora. The small value is in agreement with the results of Spitale et al. (2006). (c) Like (a) therefore the system formed by Atlas-Prometheus-Pandora has been considered. The addition of Pandora raised the diffusion coefficient in all extensions of the Corotation resonance by a factor of an order of magnitude, agreeing with the results obtained with IPS and dynamical maps.} 
	\label{fig_11}%
\end{figure}

Finally, Figure~\ref{fig_11}(c), presents the results for the complete system, considering the perturbations from both Prometheus and Pandora. In this scenario, the phase space becomes rich with resonances and complex phenomena arise. Atlas is influenced by the 54:53 resonance with Prometheus, the 70:67 resonance with Pandora, and the 121:118 resonance between Pandora and Prometheus (Renner et al. 2016).

As discussed in the previous sections, Prometheus perturbs Atlas more significantly than Pandora. Therefore, the phase space in the Atlas-Pandora scenario shows larger regions of periodic and quasi-periodic motion than in the Atlas-Prometheus scenario. 

Nonetheless, in the full system, the resonance overlap described in Section ~\ref{4.2} likely contributes to Atlas's chaotic motion. In the chaotic diffusion analysis, the region around the Corotation resonance becomes more globally chaotic in this orbital configuration: almost all initial conditions integrated have higher diffusion coefficient values $\delta_t f>10^{-3}$, while most lie around $\delta_t f\approx10^{-2}$.

None of the initial conditions reached values associated with strong chaotic motion ($\delta_t f>10^{-1}$), indicating that even with the overlap of resonances, the chaotic layer associated with the separatrix between the mean-motion resonances and the corotation zone remains confined to lower eccentricities.

This confinement has helped prevent the onset of more extreme chaos by limiting close encounters and catastrophic collisions while allowing Atlas to diffuse irregularly along the stochastic layer between the resonances. This dynamical configuration suggests the long-period stability of Atlas while confirming its local chaotic nature.

The results are consistent with Renner et al. (2016), where the overlap between the Prometheus-Atlas and Pandora-Atlas resonances increases the system's overall chaotic nature. However, the effects are more pronounced in the regions near the Corotation resonance than at Atlas's actual position. At this location, the diffusion coefficient $\delta_tf$ shows only slight changes when comparing the Prometheus-Atlas and the Atlas-Prometheus-Pandora diffusion coefficient mappings. 

In both cases, the diffusion remains confined to the chaotic layer associated with the separatrix between the resonance, preventing Atlas from drifting to higher eccentricities and avoiding close encounters and collisions. These results align with those of Pereira et al. (2024), where the authors investigated Atlas's chaotic dynamics using Lyapunov exponents. They concluded that Atlas's chaotic orbit has a short Lyapunov time but is radially confined.

\section{The dissipation in Prometheus and its migration: the past of Atlas' orbital dynamics}
\label{6}

The variations of the semi-major axis and eccentricity of Prometheus are the result of tidal effects on both Saturn and Prometheus. Since we are working with a gas giant and an icy satellite, the modeling for tidal evolution differs for each body. It is important to note that this analysis considers an isolated system, meaning that disturbances from other satellites and Saturn’s rings have been excluded. The models used to calculate the variations of these two orbital parameters are described below.

\subsection{Tidal dissipation on Saturn}
\label{6.1}

Saturn is modeled as a two-layer body, consisting of a viscoelastic core surrounded by a gaseous envelope. Each layer contributes separately to tidal dissipation. According to Ferraz-Mello et al. (2008), the equations for tidal effects on the planet (based on Darwin’s model) are presented as:

\begin{equation}\label{eq2}
    \dot a_{Saturn}=\frac{2a}{3n}\frac{9n^2 M_{P}R_{s}^5}{2M_{s}a^5} \frac{k_{2}}{Q} ( 1+ \frac{51}{4}e^2),
\end{equation}

\begin{equation}\label{eq3}
    \dot e_{Saturn}=\frac{57neM_{P}R_{s}^5}{8M_{s}a^5}\frac{k_{2}}{Q},
\end{equation}
where $M_{P}$ is the mass of Prometheus, $M_{S}$, and $R_{S}$ are the mass and radius of Saturn, $a$, $e$, and $n$ are the semi-major axis, eccentricity, and mean angular velocity of the satellite, respectively. The ratio $k_{2}/Q$ is given by the second-order Love number $k_{2}$ and the dissipation function $Q$.The next subsection describes tidal dissipation in the core and gas envelope.

\subsubsection{Dissipation in the planet's core}
\label{6.1.1}

Maxwell’s rheology describes the energy dissipation in the planet’s viscoelastic core. This is considered through the equations in Shoji and Hussmann (2017) and Remus et al. (2015)
\begin{equation}\label{(eq4)}
   \widetilde{{k}_{2s}}= \frac{3}{2} \frac{ \widetilde{\epsilon}+ \frac{2}{3} \beta}{\alpha \ \widetilde{\epsilon} - \beta},
\end{equation}
\begin{equation}\label{eq5}
     \alpha = 1 + \left(\frac{5}{2}\right) \left( \left(\frac{\rho_c}{\rho_0} \right)- 1 \right) \left( \frac{R_c }{R_S} \right)^3,
\end{equation}
\begin{equation}\label{eq6}
     \beta = \frac{3}{5} \left( \frac{R_c}{R_S} \right)^2 \left( \alpha -1 \right),
\end{equation}
\begin{equation}\label{eq7}
  \widetilde{\epsilon} = \frac { \frac{19 \widetilde{\mu}_c}{2 \rho_c g_c R_c} + \frac{\rho_0}{\rho_c} \left( 1- \frac{\rho_0 }{\rho_c} \right) \left( \beta +\frac{3}{2} \right) + \left( 1- \frac{\rho_0}{\rho_c} \right)} {\left( \alpha + \frac{3}{2} \right) \frac{\rho_0}{\rho_c} \left(1- \frac{\rho_0}{\rho_c} \right)},
\end{equation}
being $Q=|\widetilde{k_{2s}}|/ |\text{Im}[\widetilde{k_{2s}}|$, $\text{Im}[\widetilde{k_{2s}}]= k_{2s}/Q$, with $k_2= \widetilde{k_{2s}}$. The parameters $g_c$, $R_c$, $\rho_c$ are the gravitational acceleration on the surface of the core, the radius of the core, and its density. $\rho_0$ is the density of the envelope and the constants $\alpha$ and $\beta$ are depends on the physical parameters of the planet. The complex of Maxwell's frequency (Remus et al. 2012) is given by

\begin{equation}\label{eq8}
	\tilde{\mu}_{c} = \frac{w_{f}\mu_{c}\eta_{c}}{w_{f}\eta_{c}+i\mu_{c}},
\end{equation}
where $i = \sqrt{-1}$, $\mu_{c}$ and $\eta_{c}$ are the rigidity and viscosity of the core, the $w_{f} = 2(\Omega-n)$ is the frequency associated with the planet's rotation speed, $\Omega$.
Table 3 presents values for Saturn's core and envelope parameters. The other parameters were calculated.

\begin{table}[!ht]
\begin{center}
\caption{Parameter values for Saturn’s core and envelope. The Earth’s mass is given by $M_{\oplus} = 5.9736\times 10^{24}$ kg (Remus et al. 2014; Shoji and Hussmann 2017).}
	\begin{tabular}{ccc}
		\hline
         \hline
		Parameter & Symbol & Value \\
        \hline
		Mass of Saturn's core & $M_{c}(kg)$ & $18.65M_{\oplus}$ \\ 
		Radius of Saturn's core & $R_{c}(km)$ & $0.219R_{S}$ \\ 
		Saturn core density & $\rho_{c} (kg/m^{3})$ & $1156.3$ \\ 
		Density of Saturn's envelope & $\rho_{0} (kg/m^{3})$ & $503.587$ \\
		Rotation rate of Saturn & $\Omega (rad s^{-1})$ & $1.64\times10^{-4}$ \\ 
        \hline
	\end{tabular}
 \end{center}
 \label{table 3}
\end{table}

\subsubsection{Dissipation in the planet's gas envelope}
\label{6.1.2}

Tidal dissipation in the planet's envelope, due to the mechanism known as resonance locking (RL) proposed by Witte and Savonije (1999, 2001) has been analyzed.

Planets such as Jupiter and Saturn have convective envelopes, in which the oscillating fluid generates inertial waves (Greenspan et al. 1968). These inertial waves are considered attractors for tidal energy dissipation (Fuller et al. 2016). As the planet’s internal structure evolves, a satellite can become locked near a resonance and migrate. This is the RL mechanism. To describe this mechanism, we use the work of Lainey et al. (2020) where the dissipative ratio is given
\begin{equation}
	\frac{k_{2}}{Q(t)} = \frac{B}{3}\frac{M_{s}^{\frac{1}{2}}a_{0}^{\frac{1}{B}}}{G^{\frac{1}{2}}M_{p}R_{S}^{5}t_{0}}a(t)^{\left(\frac{13}{2}-\frac{1}{B}\right)},
\end{equation}
where $M_{S}$ and $R_{S}$ are the mass and radius of Saturn and $M_{P}$ is the mass of Prometheus. The parameters $a_{0}$ and $t_{0}$ are the satellite's current semi-major axis and the planet's age, while $B$ is a constant associated with the motion of the gas in the planet's envelope.

\subsection{Tidal dissipation on Prometheus}
\label{6.2}

To describe the tidal dissipation on the satellite, we use creep theory, which is similar to Maxwell’s model in rheology due to viscosity being the fundamental parameter in both (Ferraz-Mello 2013; 2015). It is worth mentioning that we model Prometheus as a homogeneous satellite that is in synchronous rotation with Saturn. Considering Folonier and Ferraz-Mello (2017) we have
 \begin{equation}\label{eq10}
     \gamma= \frac{3gM_P}{8\pi R_{S}^2 \eta_P},
 \end{equation}
 \begin{equation}\label{eq11}
    \varepsilon_{\rho} = \frac{15}{4} \frac{M_P}{M_S} \left(\frac{R_P}{a}\right)^3,
  \end{equation}
and
  \begin{equation}\label{eq12}
  \dot a_{Prometheus} = - \frac{42 R_S^2 \overline{\epsilon}_{\rho} e^2}{5a} \frac{n^2 \gamma}{n^2 + \gamma^2},
 \end{equation}
 \begin{equation}\label{eq13}
  \dot e_{Prometheus} = - \frac{21 R_S^2 \overline{\epsilon}_{\rho} e\left( 1-e^2 \right)}{5a^2} \frac{n^2 \gamma}{n^2 + \gamma^2}.  
 \end{equation}

The parameter $\gamma$ is the relaxation factor and is inversely proportional to the viscosity of the satellite. For Prometheus, we assume $\gamma = 10^{-11}s^{-1}$ (see Table 1 of Ferraz-Mello (2013)).

\subsection{Temporal variation of orbital parameters}
\label{6.3}

From the ratios between Equation~\ref{eq2} and Equation~\ref{eq12}, and Equation~\ref{eq3} and Equation~\ref{eq13}, we conclude that $\dot{a}_{Saturn}\gg\dot{a}_{Prometheus}$ and that $\dot{e}_{Saturn}\ll\dot{e}_{Prometheus}$. Therefore, the tide on the Saturn determines the temporal variation of the semi-major axis of the satellite and the tide on the satellite determines the variation of its eccentricity over time.

Rossignoli et al. (2019) estimated the surface age of Saturn's small satellites based on crater counts. The authors estimated the age of Atlas as $\tau_{at}\approx4.27\times10^{9}$ years, Prometheus as $\tau_{Pro}\approx1\times10^{9}$ years and Pandora as $\tau_{Pan}\approx2\times10^{9}$ years. 

\begin{figure}[!ht]
	\centering 
	\includegraphics[width=1.0 \columnwidth,angle=0, scale = 0.85]{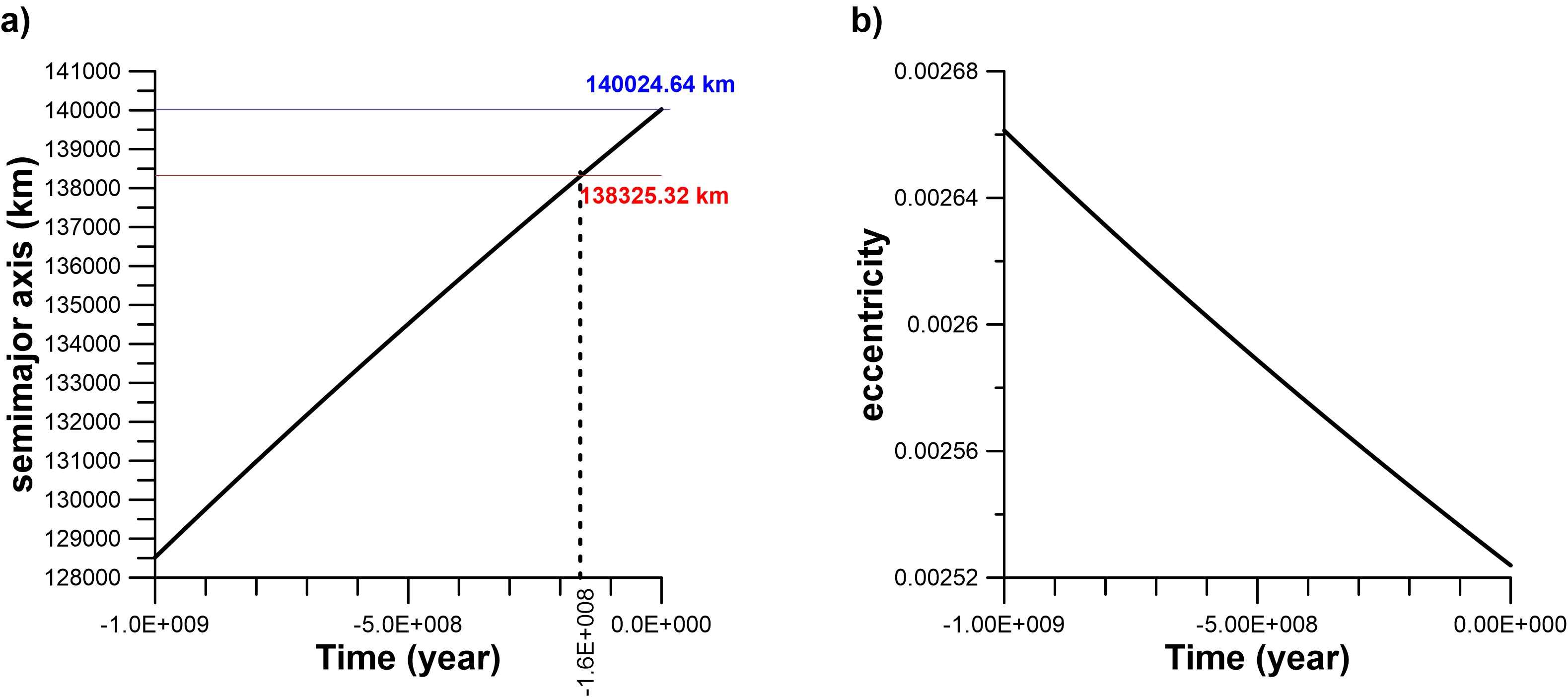}
	\caption{Variation for Prometheus (a) semi-major axis and (b) eccentricity considering the viscosity of core $10^{14}$ Pa.s and the relaxation factor $\gamma = 10^{-11}s^{-1}$ Pa.s allowing a slow variation in the Prometheus eccentricity. The values for the current semi-major axis of Atlas and Prometheus are represented by the red and blue horizontal lines respectively. We estimate that Prometheus was close to Atlas at about $-1.6\times10^{8}$ years.}
	\label{fig_12}
\end{figure}

To construct the Figure \ref{fig_12}, we used the {dissipative ratios}\footnote{Calculating the dissipative ratio $\frac{k_2}{Q}$ due to each layer we find for the core $\left[\frac{k_2}{Q}\right]_{core} \approx 3.38 \times 10^{ -5}$ and for the gas envelope $\left[\frac{k_2}{Q}\right]_{envelope} \approx 1.7 \times 10^{-3}$. These results indicate that the influence of dissipation in the gas envelope is dominant in the variation of the semi-major axis of Prometheus.} referring to each layer in Equation~\ref{eq2} and integrate it with Equation~\ref{eq13} for $10^{9}$ years. We assume $\eta_{c} = 10^{14}$Pa.s, $\mu_{c} = 10^{11}$ Pa, $\gamma = 10^{-11}s^{-1}$ and $B=1/3$.

\subsection{The dynamical map}

The effects of Prometheus's expansion on Atlas's motion were analyzed through dynamical mapping by varying the semi-major axis and the eccentricity of Prometheus. This mapping was presented by Giuppone et al. (2022) when investigating the effect of Charon's orbital expansion on Pluto's other six small satellites.

A grid of numerical simulations was constructed, considering Saturn as the main body and analyzed the expansion of Prometheus in two different scenarios i) considering only Atlas and ii) considering Atlas and Pandora. We also included Saturn's oblatenesses by accounting for the $J_{2}$, $J_{4}$ and $J_{6}$ terms of its gravitational potential.

For each initial condition, we numerically solved the exact motion equations using the MERCURY package (Chambers 1999) and selected the RA 15 algorithm (Everhart 1985). We calculated the maximum variation in Atlas's semi-major axis ($\Delta a$) following Rodríguez $\&$ Callegari (2021). The initial orbital elements for Atlas and Pandora are provided in Table~\ref{table 2}. Additionally, the initial orbital elements for Prometheus are provided in Table ~\ref{table 2}, except for the values of its semi-major axis and eccentricity,

Figure~\ref{fig_12}(a) presents the lower bound for the semi-major axis of Prometheus, which we set at $a_{Pro}=138,150$ km corresponding to an age of  $\tau_{Pro}\approx 1.6\times10^{8}$ years. As the current osculating semi-major axis of Prometheus is $\sim$140,024.64 km, we extended the upper bound to highlight the resonance near Atlas's current position.

\begin{figure}[!ht]
	\centering 
	\includegraphics[width=1.0 \columnwidth,angle=0, scale = 0.65]{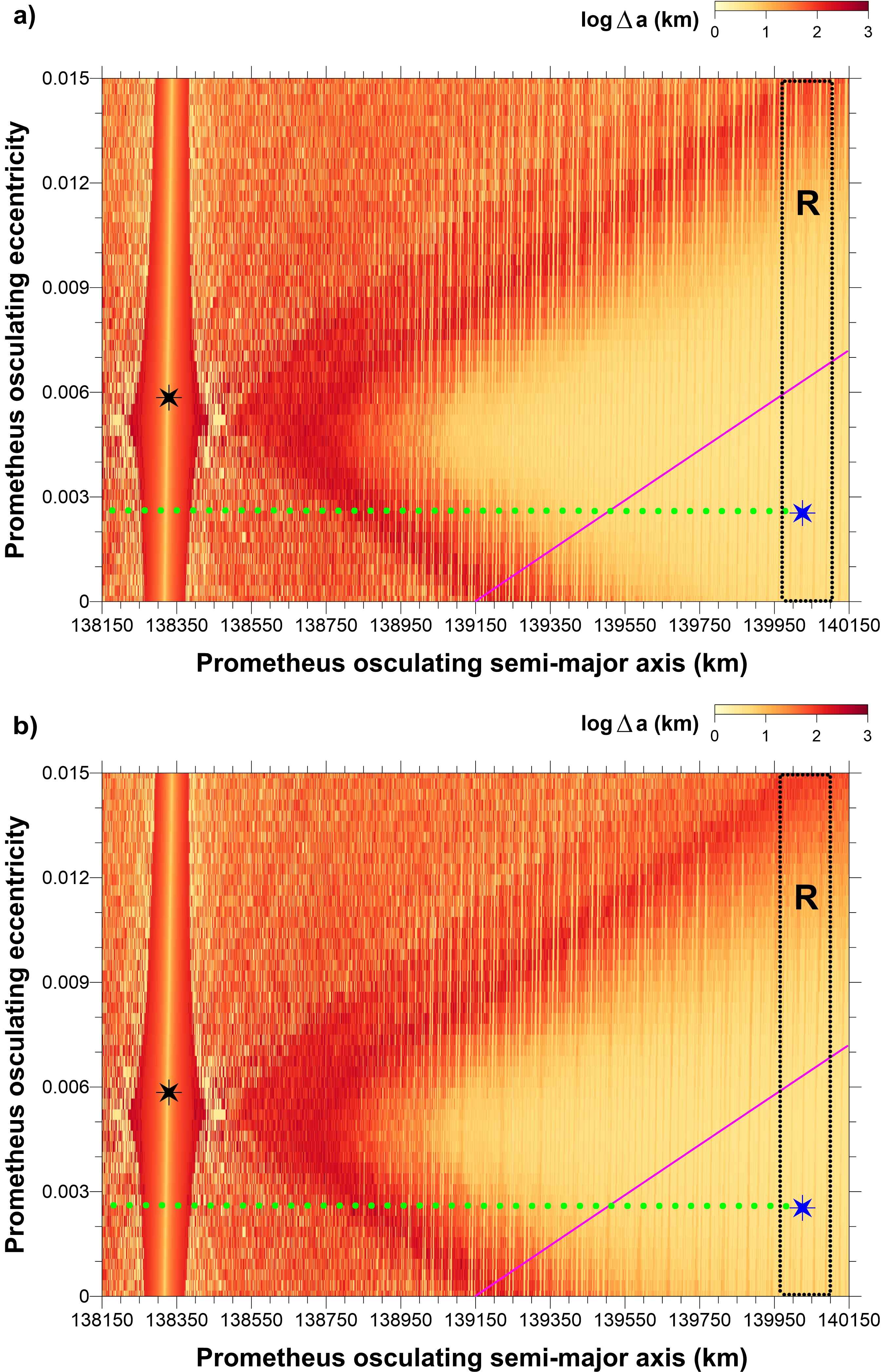}
    \caption{Dynamical map for Atlas varying initial semi-axis and eccentricity of Prometheus. The model includes (a) Prometheus and Atlas and (b) Prometheus, Atlas, and Pandora along with the Saturn potential coefficients $J_{2}$, $J_{4}$ and $J_{6}$. The initial orbital elements of Atlas are assumed to be fixed on January 01, 2000. We performed 127,755 numerical simulations integrating over 100 years with a time interval of 0.1 years. For each simulation, we calculated the maximum variation of the Atlas semi-major axis ($\Delta a$). The red and blue stars represent the current orbital position of Atlas (138,325.32 km, 0.0059) and Prometheus (140,024.64 km, 0.00252). The regions with the highest $\log\Delta a$ values correspond to the irregular motion and those with the lowest $\log\Delta a$ values correspond to the locations for the resonances between Prometheus and Atlas. The green dotted line indicates the tidal orbital evolution of Prometheus. The region for co-orbital motion is discussed in Section ~\ref{6.5}. \textbf{R} represents the region for 1st order and higher order resonances discussed in Section~\ref{6.6}. The magenta line represents the collision curve between Prometheus and Atlas.}
	\label{fig_13}
\end{figure}

By considering the interval [138,150 km, 140,150 km] for the semi-major axis and [0, 0.015] for the eccentricity of Prometheus clones, we generated the map shown in Figure~\ref{fig_13}. The black and blue stars represent the current orbital positions of Atlas and Prometheus. The dotted line depicts the Prometheus's migration. High values of $\log\Delta a$ indicate irregular motion, while low values suggest regular motion.

Both figures show that around the current position of Atlas, there is a vertical band with a width $\sim$ 100 km, delimited by borders with high value for $\log\Delta a$. Its inner region contains a narrow band with small values for $\log\Delta a$. This region represents the co-orbital resonance zone, which is discussed in detail in Section ~\ref{6.5}. In Figure~\ref{fig_13}(a) another region with a small value for $\log\Delta a$ is labeled \textbf{R} and is delimited by a dotted rectangle near the current orbit of Prometheus. This is discussed in Section ~\ref{6.6}. The addition of Pandora, as shown in Figure~\ref{fig_13}(b) does not significantly affect Atlas's orbital dynamics. This observation will be analyzed in the following sections.

\subsection{The dynamics in the past: the co-orbital case}
\label{6.5}

\begin{figure}[!h]
	\centering 
	\includegraphics[width=1.0 \columnwidth,angle=0, scale = 0.65]{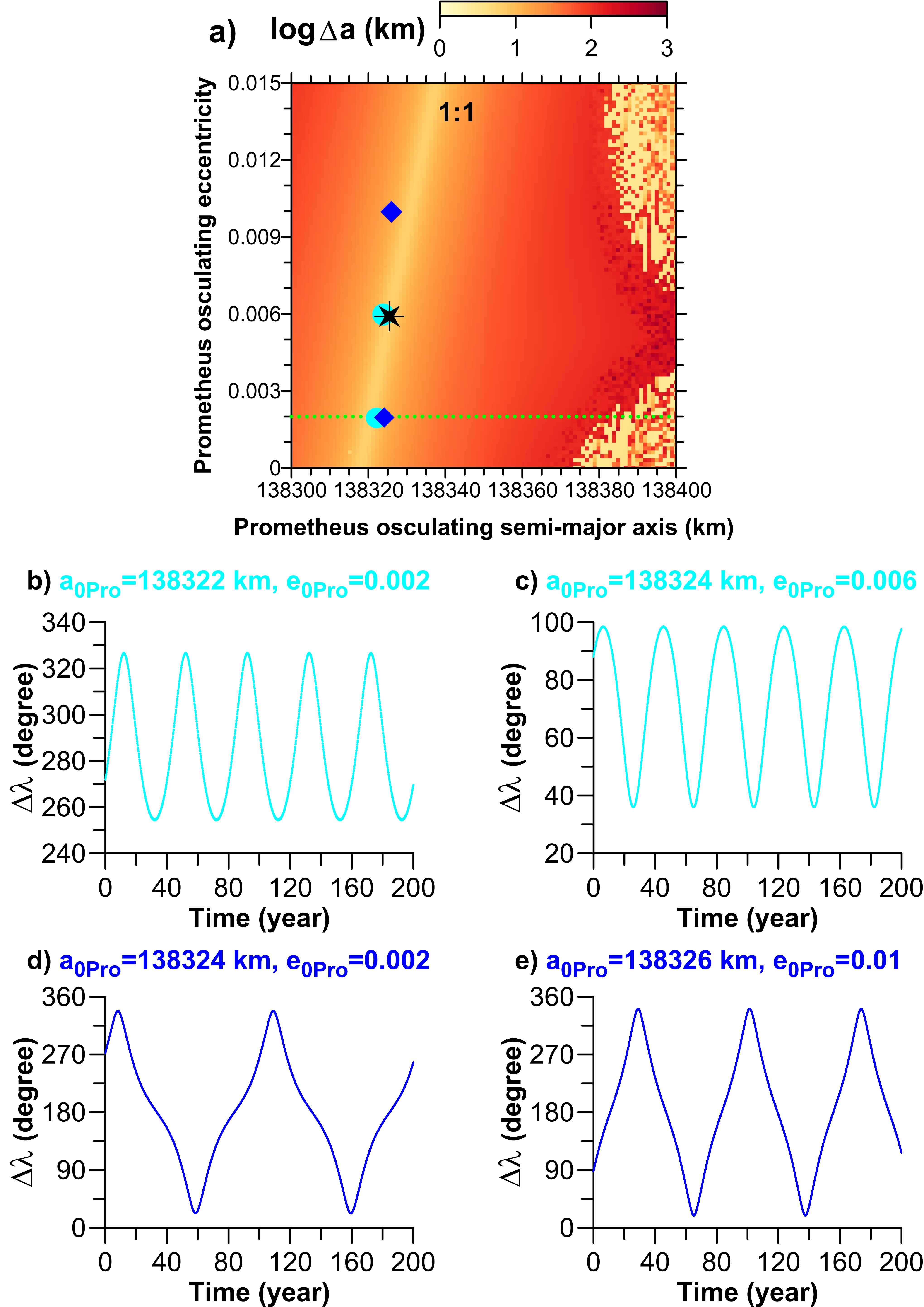}
	\caption{Dynamical map for co-orbital motion. The black star represents the current position of Atlas. The colored symbols represent the initial conditions for Prometheus clones. The green dotted curve represents the tidal evolution of Prometheus. In (a) we observe the zone for the co-orbital resonance and the lighter band represents the equilibrium solutions. (b) and (c) examples for the tadpole orbital motion and (d) and (e) horseshoe orbital motion. In (a) we performed 10,201 numerical simulations integrated over 100 years with a time interval of 0.1 days.}
	\label{fig_14}
\end{figure}

Figure~\ref{fig_14}(a) represents the phase space mapping for the 1:1 co-orbital motion in the interval [138,300 km, 139,400 km] for the semi-major axis. The inner diagonal white line represents the equilibrium solution. The black star represents the current position of Atlas and the green dotted curve represents the tidal evolution of the Prometheus clone. This scenario reveals the co-orbital motion for the Atlas-"Prometheus clone" pair. The boundaries of this region show high values for $\log\Delta a$ associated with the separatrices.

Figure~\ref{fig_14}(a) shows the mapping for the Prometheus-Atlas scenario and Figures~\ref{fig_14}(b-e) represents the temporal variation for the argument $\Delta \lambda$, which represents the difference between the mean longitudes of Atlas and Prometheus clone for four initial conditions given by colored symbols in Figure~\ref{fig_14}(a).

The cyan circles in Figure~\ref{fig_14}(a) represent initial conditions in which the orbit found follow the tadpole regime (Figures~\ref{fig_14}(b,c)), and the blue diamonds in Figure~\ref{fig_14}(a) represent initial conditions with horseshoe orbits (Figures~\ref{fig_14}(d,e)).

\subsection{The dynamics in the past: the vicinity around the current orbit of Prometheus}
\label{6.6}

The initial semi-major axis of Prometheus, ranging from 139,990 km to 140,100 km was adopted for constructing the dynamical map representing region \textbf{R} in Figure~\ref{fig_13}(a).

\begin{figure}[!ht]
	\centering 
	\includegraphics[width=1.0 \columnwidth,angle=0, scale = 0.85]{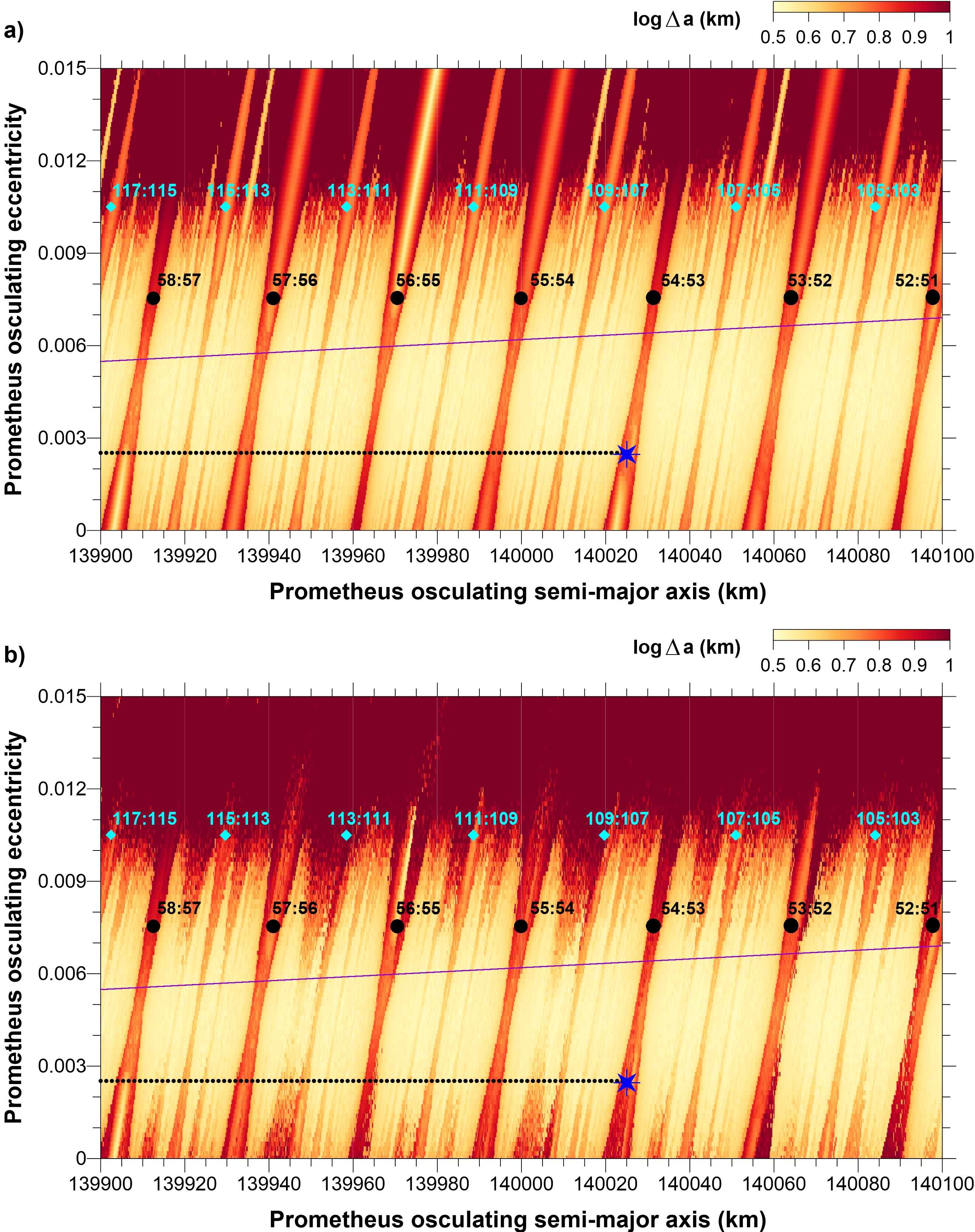}
	\caption{Same as Figure~\ref{fig_13}, therefore the \textbf{R} region that represents the vicinity for the current orbit of Prometheus, represented by the blue star had been considered. The black dotted line indicates the tidal orbital evolution of Prometheus, and the magenta line represents the collision curve between Prometheus and Atlas. Several 1st order resonances are crossed along the Prometheus migration. (a) represents the scenario with Atlas and Prometheus and (b) Pandora in this scenario is included. Note the inclusion of Pandora affects higher-order resonances.}
	\label{fig_15}
\end{figure}

As expected, in Figures~\ref{fig_15}(a,b) the 54:53 Corotation resonance between Prometheus and Atlas at Prometheus's current position (blue star) is identified. Additionally, several other mean-motion resonances (MMRs) between both satellites are seen, with the strongest being the first-order resonances 58:57, 57:56, 56:55, 55:54, and 54:53. Higher-order weak resonances are also present. All these first-order MMRs are of the Corotation type, where the oscillating critical angles have the general form $\phi_{2}^{m+1:m} = (m+1)\lambda-m\lambda_{At}-\varpi$, where $m$=53, 54, 55, 56, 57 and  $\lambda$, $\lambda_{At}$ and $\varpi$ represent the mean longitude of the Prometheus clone, the mean longitude of the Atlas, $\varpi$ is the longitude of the pericenter of the Prometheus clone. These angles oscillate around 180°, indicating that the conjunctions occurred near the apocenter of the Prometheus clone. For second-order resonances, the domain in the phase space is visible. However, this does not apply to all third-order resonances, making it difficult to identify their centers.

The collision curve between Prometheus and Atlas is represented by the magenta line in Figures\ref{fig_15}(a,b). and Prometheus's tidal orbital evolution is shown by the black dotted line in Figures~\ref{fig_15}(a,b), highlighting the crossing of MMRs with Atlas during orbital expansion. When Prometheus crosses any of the first-order MMRs shown in Figures~\ref{fig_15}(a,b), the semi-major axis of Atlas does not undergo significant variations (on the order of 10 km), except for high eccentricities of the Prometheus clone.

Comparing the collision curves with the expansion, we see that the eccentricity of Prometheus remained close its current value ($\sim$0.00252), much lower than the value found for the collision.

The inclusion of Pandora contributes to changes in the resonant structures, as seen in Section ~\ref{4.2}. There is an overlap between the resonances caused by Prometheus and Pandora, with Figure~\ref{fig_15}(b) showing that higher-order resonances are the most affected.

\subsection{The probability of capture}

El Moutamid et al. (2017) provided analytical expressions to determine the probability of capture in first-order mean-motion resonances where a migrating satellite is disturbing a non-migrate particle. Denoting the width of the resonance $W_{CER}$ (Equation (2) from that article), the probability of capture is given by $P = W_{CER}/\pi a_{0}$ (its Equation (22)). $a_{0}$ is the center of the resonance, which we assume to be $\sim$ 138,323.25 km, the approximate value for the center of the region \textbf{C} shown in Figure~\ref{fig_12}. Table ~\ref{table 4} shows the $W_{CER}$ and the capture probabilities of past resonances with Prometheus shown in Figure~\ref{fig_15}.

\begin{table}[!ht]
\caption{Probabilities of capture and the width of the corresponding $W_{CER}$ Corotation mean-motion resonances shown in Figure~\ref{fig_15}. The values $a_{0}$=138,323.5 km and $e_{S}$=0.0025 were used to determine the commensurability. We use Equations (2) and (22) given in El Moutamid et al. (2017).}
	\centering
	\begin{tabular}{cccc}
		\hline
        \hline
		Resonance & $a_{S}$(km) & $W_{CER}$ (km) & Probability ($\%$) \\ \hline
		58:57 & 139,900 & 3.63 & 0.00084 \\
		57:56 & 139,934 & 3.599 & 0.00083 \\
		56:55 & 139,971 & 3.567 & 0.00082 \\
		55:54 & 139,990 & 3.534 & 0.00081 \\
		54:53 & 140,024.65 & 3.515 & 0.0008 \\ 
        \hline
	\end{tabular}
 \label{table 4}
	
\end{table}

The results in Table ~\ref{table 4} show that if the orbit of the disturbance is external to the disturbed body, there is an increase in the probability of capture due to Prometheus's semi-major axis. A similar trend occurs with the width of $W_{CER}$, indicating that the larger the Corotation domain, the greater the trapping capture at a given resonance.

\section{Conclusions}
\label{7}

Small satellites experience secular and resonant disturbances caused by larger satellites, which affect their semi-major axis, eccentricities, and orbital inclinations. In the case of resonant perturbations, understanding the orbital dynamics of a small satellite is essential. Previous studies on Anthe, Methone, and Aegaeon, contributed to understanding these phenomena and served as a basis for applying them to the Atlas (e.g. Callegari and Yokoyama 2020; Callegari et al. 2021; Callegari and Rodríguez 2023).

Mapping the phase space for the 54:53 MMR with Prometheus revealed that Atlas is located at the boundary between the separatrices of the Corotation resonance. This position is responsible for the non-periodic time variations of the geometric arguments $\phi_{2}$ and $\phi_{1}$. Similarly, the map for the 70:67 MMR with Pandora showed that Atlas's orbital location is also outside this resonance. The widths of the 54:53 and 70:67 resonances in frequency space are approximately 4 km and 2 km, respectively, leading to a very low capture probability (see Table~\ref{table 4}).

By mapping the 54:53 MMR, we identified the regions where the Corotation and Lindblad resonances dominate. The Corotation region is limited to orbital eccentricity values between 0.003 $\leq$ $e_0$ $\leq$ 0.006 and occupies a narrow band in phase space $\sim$ 4 km within the semi-major axis range of 138,321 km $\leq$ $a_0$ $\leq$ 138,325 km. Atlas clones with orbits inside this region experience liberation of the geometric argument $\phi_{2} = 54\lambda_{Pro}-53\lambda_{S}-\varpi_{Pro}$ around 180°.

For initial eccentricity of $e_0$ $\geq$ 0.01 and initial semi-major axis $a_0$ $\geq$ 138,327 km, the Lindblad resonance domain was identified. For clones inside this region, the geometric argument $\phi_{1} = 54\lambda_{Pro}-53\lambda_{S}-\varpi_{S}$ oscilates around 0.

Expanding the interval for the mapping given by Ceccatto et al. (2022), revealed two first-order resonances, 53:52 to the left and 55:54 to the right of the Atlas's current orbit. For each region, we identified the initial conditions ($a_0$, $e_0$) for Atlas clones where oscillations of arguments related to these resonances occur (see Figure~\ref{fig_2}).

After investigating Atlas's orbital vicinity, we examined how Prometheus's orbital eccentricity contributes to Atlas's trapping within the Corotation resonance. One condition for libration of the geometric argument $\phi_{2}$, as described by Renner et al. (2016), is that Prometheus's orbital eccentricity must be close to zero. To illustrate this, we mapped a Prometheus clone with near-zero eccentricity, showing an expansion of the  Corotation region and a narrowing of the Lindblad region. However, for eccentricity values close to the current ones, the Corotation region is filled with the separatrices of the resonance (see Figure ~\ref{fig_7}).

On the other hand, mapping the phase space of Atlas's orbital vicinity with Pandora as the main perturber revealed two additional resonances, 64:67 and 73:70. Analysis of its IPS revealed the locations for the $\Psi_{4}$, $\Psi_{3}$, $\Psi_{2}$ and $\Psi_{1}$ multiples related to the 70:67 resonance, showing that Atlas's orbit is close to the $\Psi_{1}$ multiplet (see Figure~\ref{fig_8} and Figure~\ref{fig_9}).

Resonances overlap occurs when the separatrices of one resonance intersect those of another, as described by Wisdom (1980). An example is shown in Figure~\ref{fig_10} where the separatrices of the 54:53 and 70:67 resonances overlap. However, this overlap does not substantially affect the stability of the region around Atlas, keeping its orbit confined between the resonance region and the separatrices. 
This result supports the findings of Pereira et al. (2024), who concluded that Atlas's orbit chaotic only in the angular components, and not in the radial one, thus being chaotically confined. Despite the short Lyapunov times observed by Cooper et al. (2015), further dynamical analysis of the phase space around Atlas's current position is necessary, which has been conducted in this work (see Section~\ref{5}).
%This result is in agreement with the calculation of the Laskar diffusion coefficient (weak chaos). However, despite obtaining an agreement between the Lyapunov Time of the radial component and the Instability Time, Pereira et al. (2024) analyzed only two orbits over long periods of time, while the current position of Atlas on the border between the separatrices suggests the need for a more comprehensive dynamical analysis of the phase space around Atlas' current position.

%However, this overlap does not substantially alter the stability of the region in which Atlas is located, keeping the orbit confined to the region of the separatrices between the resonances in regions of low eccentricity. The finding that Atlas's orbit is weakly chaotic (or confined chaotic) is in agreement with Pereira et al. (2024), in which the authors estimated, using Lyapunov exponents, that although Atlas's orbit is chaotic, it does not present large variations and can be considered stable.

The study of Atlas's dynamic past revealed that during Prometheus's tidal migration, Atlas and Prometheus shared co-orbital dynamics in the recent past. The mapping analysis indicated conditions where horseshoe or tadpole orbits could occur. Further investigation of Atlas's orbital vicinity revealed the 1st- and 2nd-order resonances that were crossed in the past, with the 1st-order resonance being the strongest and having widths of approximately 5 km. The resonant structures of the Prometheus-Atlas system resemble an “hourglass”, with the center occupied by the Corotation resonance domain and the upper part by the Lindblad resonance. However, the capture probability for these first-order resonances is small, on the order of $10^{-7}$.

\section{Acknowledgements}
\label{Thanks}

We are especially grateful to two anonymous reviewers, who have reviewed this work. We also thank Adrián Rodriguez for the discussions and suggestions that contributed to the preparation of this work.

\section{Bibliographic references}
\label{bibliographic}

%%\bibliographystyle{elsarticle-harv} 
%%\bibliography{example}



%%\bibitem[Author(year)]{label}
%% For example:

%%\bibitem[Aladro et al.(2015)]{Aladro15} Aladro, R., Martín, S., Riquelme, D., et al. 2015, \aas, 579, A101

BROUWER, D. and CLEMENCE, G.M. Methods of Celestial Mechanics (Academic Press, New York) (1961).

CALLEGARI JR., N., RODRÍGUEZ, A., CECCATTO, D. T. The current orbit of Methone (S/2004 S 1). Celest. Mech. Dyn. Astr. 133, 49pp. (2021).

CALLEGARI JR., N., YOKOYAMA, T. Dynamics of Enceladus and Dione inside the 2/1 Mean- Motion Resonance under tidal disspation. Celest. Mech. Dyn. Astr., 102, 273-296 (2008).

CALLEGARI JR., N., YOKOYAMA, T. Dynamics of the 11:10 Corotation and Lindblad Resonances with Mimas, and Application to Anthe, Icarus, 348, 113820 (2020).

CALLEGARI JR., N., YOKOYAMA, T. Dynamics of Two Satellites in the 2:1 Mean-Motion Resonance: application to the case of Enceladus and Dione. Celest. Mech. Dyn. Astr. 98, 5-30 (2007).

CALLEGARI JR., N., YOKOYAMA, T. Numerical exploration of resonant dynamics in the system of Saturnian inner Satellites. Planetary and Space Science 58, 1906-1921 (2010a).

CALLEGARI JR., N., YOKOYAMA, T. Long-term dynamics of Methone, Anthe and Pallene. Icy Bodies of the Solar System, Proceedings of the International Astronomical Union, IAU Symposium, 263, 161-166 (2010b).

CECCATTO, D. T., CALLEGARI JR., N., RODRÍGUEZ, A. The current orbit of Atlas (S XV). Proceedings of International Astronomical Union, IAU Symposium, 364, 120-127 (2022).

CHAMBERS, J. E. A hybrid sympletic integrator that permits close encounters between massive bodies. Montly Notices of the Royal Astron. Society 304, 793-799 (1999).

COOPER N. J.; RENNER S.; MURRAY C. D.; EVANS M. W. Saturn’s inner satellites orbits, e the chaotic motion of Atlas from new Cassini imaging observations. The Astronomical Journal 149, 27-45, (2015).
DANBY, J. M. A. Fundamentals of celestial mechanics. Willmann-Bell, Inc. (1988).

EL MOUTAMID M.; SICARDY B.; RENNER S. Derivation of capture probabilities for the corotation eccentric mean motion resonances, Monthly Notices of the Royal Astronomical Society, Volume 469, Issue 2, August 2017, Pages 2380–2386

EL MOUTAMID, M.; RENNER, S.; SICARDY, B.. Coupling between corotation and Lindblad resonances in the elliptic planar three-body problem. Celest. Mech. Dyn. Astr., 118, 235-252 (2014).

EVERHART, E. An efficient integrator that uses Gauss-Radau spacings. In: IAU Coloquium 83, 185-202 (1985).

FERRAZ-MELLO,  S.,  RODRÍGUEZ,  A.,  \&  HUSSMANN,  H. Celestial  Mechanics  and  Dynamical Astronomy, 101, 171, (2008).

FERRAZ-MELLO, S. Estimation of periods from unequally spaced observations. The Astronomical Journal, 36, 619-624 (1981).

FERRAZ-MELLO, S. First-order resonances in satellites orbits. In: S. Ferraz-Mello and W. Sessin (eds) Resonances in the Motion of the Planets, Satellites and Asteroids, IAG/USP, São Paulo, 37-52 (1985).

FERRAZ-MELLO, S., A\&A, 579, A97 (2015).

FERRAZ-MELLO, S., Celestial Mechanics and Dynamical Astronomy, 116, 109 (2013).

FOLONIER, H. A., \& FERRAZ-MELLO, S. Celestial Mechanics and Dynamical Astronomy, 129, 359, (2017).

FULLER, J., LUAN, J., \& QUATAERT, E. 2016, in AAS/Division for Planetary Sciences Meeting Abstracts, Vol. 48, AAS/Division for Planetary Sciences Meeting Abstracts \#48, 401.01

GIORGINI, J. D., YEOMANS, D. K., CHAMBERLIN, A. B., CHODAS, P. W., JACOBSON, R. A., KEESEY, M. S., LIESKE, J. H., OSTRO, S. J., STANDISH, E. M., WIMBERLY, R. N. JPL's On-Line Solar System Data Service. American Astronomical Society, DPS meeting \#28, id.25.04; Bulletin of the American Astronomical Society, Vol. 28, p.1158 (1996).

GIUPPONE, C. A., BEAUGÉ, C. MICHTCHENKO, T. A. S. FERRAZ-MELLO, Dynamics of two planets in co-orbital motion, Monthly Notices of the Royal Astronomical Society, Volume 407, Issue 1, September 2010, Pages 390–398.

GIUPPONE, C. A., RODRÍGUEZ, A. MICHTCHENKO, T. A., ALMEIDA, A. A. Past and present of the Circumbinary moons in the Pluto-Charon system. Astronomy \& Astrophysics, 658, 18 pp. (2022). 

GOLDREICH, P., RAPPAPORT, N. Origin of chaos in the Prometheus-Pandora system. Icarus, 166, 320-327 (2003a).

GOLDREICH, P., RAPPAPORT, N. Chaotic motions of Prometheus and Pandora. Icarus, 166, 391-399 (2003b).

GREENBERG, R. Apsidal Precession of Orbits about an Oblate Planet. The Astronomical Journal 86, 912-914 (1981).

GREENSPAN, H.,  BATCHELOR,  C.,  \&  ISERLES,  U.  1968, The Theory of Rotating Fluids, Cambridge Monographs on Mechanics (Cambridge University Press). https: //books.google.com.br/books?id=2R47AAAAIAAJ

GUIMARÃES, G. T., MICHTCHENCKO, T. A. Chaotic diffusion in the action and frequency domains: estimate of instability times. Eur. Phys. J. Spec. Top. 232, 3147–3154 (2023)

HEDMAN, M. M., COOPER, N. J., MURRAY, C. D., BEURLE, K., EVANS, M. W., TISCARENO, M. S., BURNS, J. A. Aegaeon (Saturn LIII), a G-ring object. Icarus 207, 433-447 (2010).

HEDMAN, M. M., MURRAY, C. D., COOPER, N. J., TISCARENO, M. S., BEURLE, K., EVANS, M. W., BURNS, J. A. Three tenous rings/arcs for three tiny moons. Icarus 199, 378-386 (2009).

JACOBSON, R. A., ANTREASIAN, P. G., BORDI, J. J., CRIDDLE, K. E., IONASESCU, R., JONES, J. B., MACKENZIE, R. A., MEEK, M. C., PARCHER, D., PELLETIER, F. J., OWEN JR., W. M., ROTH, D. C., ROUNDHILL, I. M., STAUCH, J. R. The Gravity Field of the Saturnian System from satellite observations and spacecraft tracking data. The Astronomical Journal 132, 2520-2526 (2006b).

JACOBSON, R. A., SPITALE, J., PORCO C. C., OWEN JR., W. M. The GM values of Mimas and Tethys and the libration of Methone. The Astronomical Journal 132, 711-713 (2006a).

JACOBSON, R. A., SPITALE, J., PORCO C. C., BEUELE, K., COOPER, N. J., EVANS, M. W., MURRAY, C. D. Revised orbits of Saturn’s inner satellites. The Astronomical Journal 135, 261-263 (2008).

LAINEY, V., CASAJUS, L. G., FULLER, J., et al. 2020, Nature Astronomy, 4, 1053, doi: 10.1038/s41550-020-1120-5

LASKAR, J. The chaotic motion of the solar system: A numerical estimate of the size of the chaotic zones, Icarus, vol. 88, p. 266 (1990)

LASKAR, J. Frequency analysis for multi- dimensional systems: global dynamics and diffusion. Physica D: Nonlinear Phenomena 67, 257–281 (1993).

MICHTCHENKO, T. A., FERRAZ-MELLO, S. Modeling the 5:2 Mean-Motion Resonance in the Jupiter-Saturn Planetary System. Icarus 149, 357-374 (2001a).

MICHTCHENKO, T. A., FERRAZ-MELLO, S. Resonant Structure of the outer solar system in the vicinity of the planets. The Astronomical Journal 122, 474-481 (2001b).

MICHTCHENKO, T. A., NESVORNÝ, D. Wavelet analysis of the asteroidal resonant motion. Astronomy and Astrophysics 313, 674 (1996).

MUNÕZ-GUTIÉRREZ, M. A.; GIULIATTI WINTER, S. Long-term evolution and stability of Saturnian small satellites: Aegaeon, Methone, Anthe and Pallene. Monthly Notices of the Royal Astronomical Society 470, 3750-3764 (2017).

MURRAY, C. D., DERMOTT, S. F. Solar System Dynamics, Cambridge University Press (1999).

PEALE, S. J. Origin and Evolution of the Natural Satellites. Annual Review of Astron. and Astrophys. 37, 533-602 (1999).
PORCO, C. C. S/2004 S 1 and S/2004 S 2. IAU Circ. 8401 (2004 August 16) (2004).

PEREIRA, L. S.; MOURÃO, D.; WINTER, O. C. Confined chaos and the chaotic angular motion of Atlas, a Saturn’s inner satellite, Monthly Notices of the Royal Astronomical Society, Volume 529, Issue 2, April 2024, Pages 1012–1018

PORCO, C. C., BAKER, E., BARBARA, J., BEURLE, K., BRAHIC, A., BURNS, J. A., CHARNOZ, S., COOPER, N., DAWSON, D. D., DEL GENIO, A. D., DENK, T., DONES, L., DYUDINA, U., EVANS, M. W., GIESE, B., GRAZIER, K., HELFENSTEIN, P., INGERSOLL, A. P., JACOBSON, R. A., JOHNSON, T. V., MCEWEN, A., MURRAY, C. D., NEUKUM, G., OWEN, W. M., PERRY, J., ROATSCH, T., SPITALE, J., SQUYRES, S., THOMAS, P., TISCARENO, M., TURTLE, E., VASAVADA, A. R., VEVERKA, J., WAGNER, R., WEST, R.. Cassini Imaging Science: Initial Results on Saturn’s Rings and Small Satellites. Science 307 1226-1236 (2005).

PRESS, W. H., TEUKOLSKY, S. A., VETTERLING, W. T., B. P. FLANNERY. Numerical Recipes in Fortran 77. Cambridge University Press (1996).

REMUS, F., MATHIS, S., ZAHN, J. P., \& LAINEY, V. 2015, AAP, 573, A23. (2015).

REMUS, F., MATHIS, S., ZAHN, J. P., \& LAINEY, V. A\&A, 541, A165, (2012).

REMUS, F., MATHIS, S., ZAHN, J.-P., \& LAINEY, V., in -, Vol. 293, Formation, Detection, and Characterization of Extrasolar Habitable Planets, ed. N. Haghighipour, 362–368 (2014).

RENNER S.; COOPER J, N.; EL MOUTMAID M.; SICARDY B.; VIENNE A.; MORRAY C. D.; SAILLENFEST M. Origin of the chaotic motion of the Saturnian satellite Atlas. The Astronomical Journal, 151, 9pp. (2016).

RENNER, S., SICARDY, B. Use of the Geometric Elements in Numerical Simulations. Celestial Mechanics and Dynamical Astronomy 94, 237-248 (2006).

RODRÍGUEZ A., CORREA-OTTO J. A., MICHTCHENKO T. A. Primordial migration of co-orbital satellites as a mechanism for the horseshoe orbit of Janus–Epimetheus, Monthly Notices of the Royal Astronomical Society, Volume 487, Issue 2, August 2019, Pages 1973–1979

RODRÍGUEZ, A., CALLEGARI JR., N. Dynamical stability in the vicinity of Saturnian small moons. The cases of Aegaeon, Methone, Anthe and Pallene. Monthly Notices of the Royal Astronomical Society 506, 5093-5107 (2021).

ROSSIGNOLI N. L., Di SISTO R. P., ZANARDI M., DUGARO A. Cratering and age of the small Saturnian satellites, A\&A, 627, A12, 2019.

SHOJI, D., \& HUSSMANN, H. AAP, 599, L10, (2017).

SIQUEIRA, P. B. Adaptação do integrador Rebound para o estudo de anéis planetários. 116 f. Dissertação (Mestrado) - Universidade Estadual Paulista, Faculdade de Engenharia de Guaratinguetá, 2019.

SPITALE, J. N., JACOBSON, R. A., PORCO, C. C., OWEN, JR, W. M. The Orbits of Saturn’s Small Satellites Derived from Combined Historic and Cassini Imaging Observations. The Astronomical Journal 132, 792-810 (2006).

THOMAS, P. C., HELFENSTEIN, P. The small inner satellites of Saturn: Shapes, structures and some implications. Icarus 344, 113355 (2020).

THOMAS, P.C., BURNS, J.A., HEDMAN, M., HELFENSTEIN, P., MORRISON, S., TISCARENO, M.S., VEVERKA, J. The inner small satellites of Saturn: a variety of worlds. Icarus 226, 999–1019 (2013).

WACHLIN, F. C., FERRAZ-MELLO, S. Frequency map analysis of the orbital structure in elliptical galaxies. Monthly Notices of the Royal Astronomical Society 298 (1), 22– 32 (1998).

WISDON, J. The resonance overlap criterion and the onset of stochastic behavior in the restricted three-body problem. Astronomical Journal, vol. 85, Aug. 1980, p. 1122-1133.

WITTE, M. G., \& SAVONIJE, G. J. A\&A, 350, 129. (1999).

%%\end{thebibliography}

%\appendix

\appendix

\section{The influence of Pandora's mass on Atlas dynamics}
\label{A2}

Although Pandora does not contribute to significant perturbations in Atlas orbital dynamics, the combined effects between the Prometheus-Pandora pair (overlap between resonances, Section~\ref{4.2}) contribute to the emergence of stable chaos in its orbit.

\begin{figure}[!ht]
	\centering 
	\includegraphics[width=1.0 \columnwidth,angle=0, scale = 0.80]{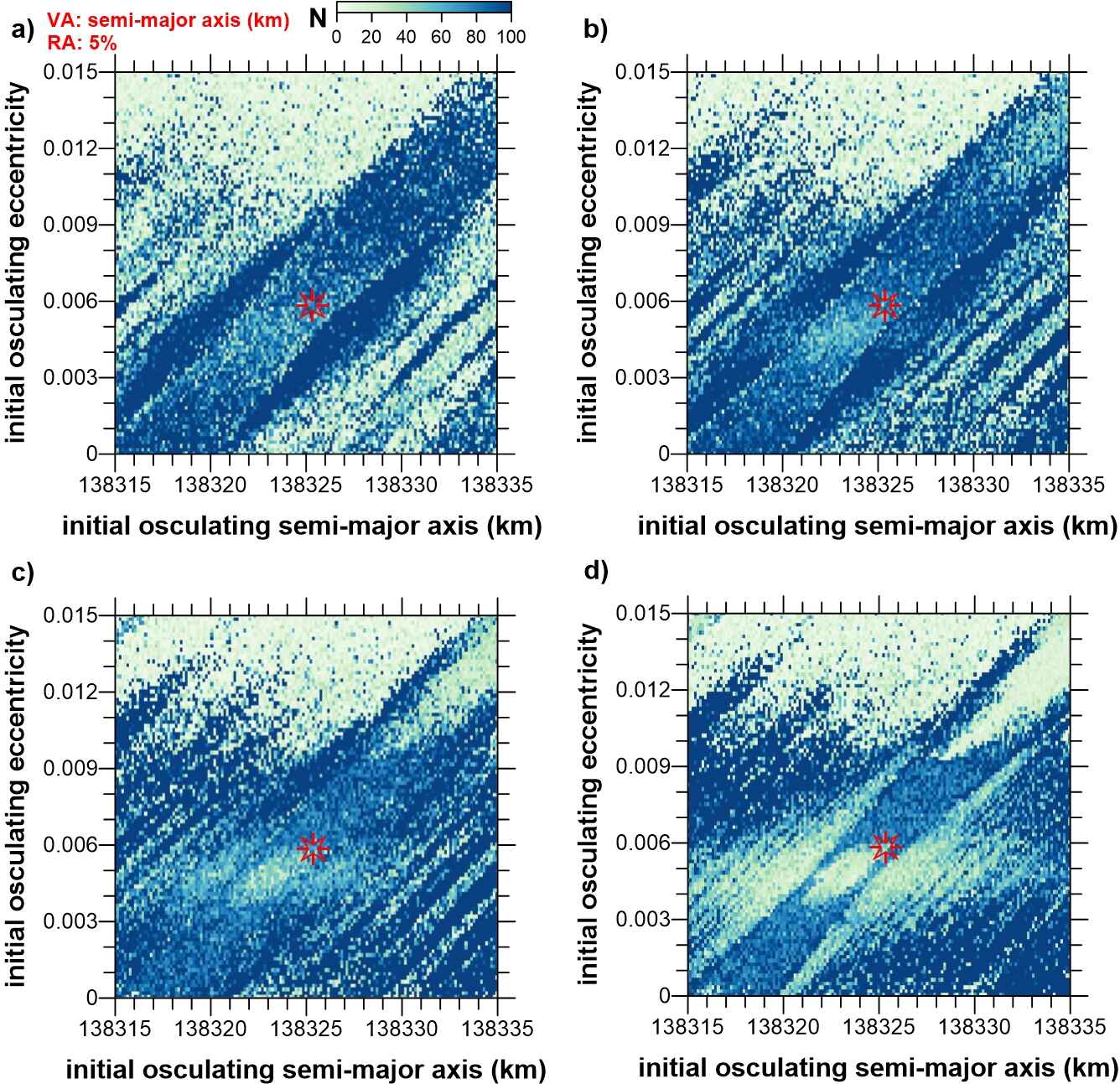}
	\caption{Mapping to the phase space of the orbital vicinity of Atlas as Figure~\ref{fig_1}(b) considering the effects of the disturbance caused by Pandora clones that vary their masses: (a) $M_{Pandora}$, (b) 0.5$M_{Pandora}$, (c) 0.25$M_{Pandora}$ and in (d) 0.1$M_{Pandora}$. We observed how the gradual decrease in Pandora's mass contributed to the emergence of the domain of Corotation and Lindblad resonances.}
	\label{fig_A1}
\end{figure}

To analyze how these effects change the mapping of the phase space, several mappings considering, in addition to Prometheus, the effects of a Pandora clone varying its mass, and the result can be seen in Figure~\ref{fig_A1} had been carried out. 

In Figure~\ref{fig_A1}(a) we have a zoom of Figure~\ref{fig_10}(c), in Figure~\ref{fig_A1}(b-d). A clone of Pandora with a mass equivalent to 0.5$M_{Pandora}$, 0.25$M_{Pandora}$ and 0.1$M_{Pandora}$, respectively, has been considered.

It can observed that with the decrease in masses, the domains related to Corotation and Lindbald resonances become more evident in phase space.

\end{document}